\newcommand{\isEmbedded}{true}
\documentclass[11pt]{article}
\usepackage{amsmath, amsthm, amssymb}
\usepackage{latexsym}
\usepackage{subfig}
\usepackage{graphicx}
\usepackage{wrapfig}
\usepackage{verbatim}
\usepackage{chapterfolder}
\usepackage{stmaryrd}
\usepackage{algorithm}
\usepackage[noend]{algpseudocode}
\usepackage{pdfpages}

\usepackage[pdftex,hypertexnames=false,linktocpage=true]{hyperref} 
\hypersetup{colorlinks=true,linkcolor=black,anchorcolor=black, citecolor=black}

\newcommand{\OR}{{\sf OR} }
\newcommand{\ORi}{\mbox{$\mathsf{OR}$} }
\newcommand{\SR}{{\sf SR}}
\newcommand{\SRi}{\mbox{$\mathsf{SR}$}}

\newcommand{\occurred}[1]{{\sf occurred}(#1)}
\newcommand{\tocc}[2]{{\sf occurred}_{#1}(#2)}

\newcommand{\Proc}{{\mathbb{P}}}
\newcommand{\Lang}{\ensuremath{\cal L}}
\newcommand{\vphi}{\ensuremath{\varphi}}
\newcommand{\sat}{\vDash}
\newcommand{\nsat}{\nvDash}
\newcommand{\angles}[1]{\ensuremath{\langle{#1}\rangle}}

\newcommand{\bound}[1]{max_{#1}}
\newcommand{\bij}{\bound{i\!j}}
\newcommand{\Rrep}{{\cal R}}

\newcommand{\fut}{\mathsf{fut}}
\newcommand{\pas}{\mathsf{past}}

\newcommand{\tspc}{\dashrightarrow}
\newcommand{\spc}{\rightsquigarrow}

\newcommand{\gammax}{\gamma^{\mathsf{max}}}

\newcommand{\fip}{\mathsf{fip}}

\newcommand{\imp}{\Rightarrow}

\newcommand{\necaff}{\mathsf{\Box aff}}

\newcommand{\necunaff}{\mathsf{\Box unaff}}

\newcommand{\node}[1]{\ensuremath{{\angles{#1}}}}
\newcommand{\CV}{{\mathbb{V}}}
\newcommand{\rsp}[1]{\mbox{\boldmath${#1}$}}

\newcommand{\toto}{xxx}
\newenvironment{proofT}{\noindent{\bf Proof }}
{\hspace*{\fill}$\Box_{Theorem~\ref{\toto}}$\par\vspace{3mm}}
\newenvironment{proofL}{\noindent{\bf Proof }}
{\hspace*{\fill}$\blacksquare$\par\vspace{3mm}}

\newtheorem{definition}{Definition}
\newtheorem{theorem}{Theorem}
\newtheorem{lemma}{Lemma}

\newtheorem*{rtheorem}{{\bf Theorem}~\ref{\toto}}
\newtheorem*{rlemma}{{\bf Lemma}~\ref{\toto}}

\newcommand{\only}[2]{#2}
\newcommand{\uncover}[2]{#2}

\newif\iflazytikz 
\lazytikztrue 
\iflazytikz
	\usepackage{graphicx} 
	\usepackage{tikzexternal}
\else
	\usepackage{tikz}
	\ifx \isEmbedded \undefined
\usetikzlibrary{3d,arrows,positioning,decorations.pathmorphing,shadows, patterns,
						backgrounds,shapes, external, intersections, calc, lindenmayersystems}

\tikzset
{
    timeline/.style =
    {
        ->,
        line width=0.5pt,
        color=black
    },
    agent_time/.style=
    {
        shading=ball,
        ball color=gray,
        ellipse,
        minimum height=15pt,
        thin,
        draw=black
    },
     syncausality/.style=
    {
        ->,
        >=stealth,
        very thick,
        decorate,
        decoration={snake,amplitude=.6mm,segment length=3mm,post length=2mm, pre length = 2mm}
    },
     guarantee/.style=
    {
        ->,
        dashed,
        >=stealth,
        very thick
    },
     msg/.style=
    {
        ->,
        >=stealth,
        very thick
    },
    channel/.style=
    {
    very thick, bend angle=20,shorten <=1pt, shorten >=1pt
    },
    time_arrow/.style=
    {
    	->,
    	>=triangle 60,
        line width=0.2pt,
        color=black!
    },
    time_mark/.style=
    {
        black,
        anchor=base,
        yshift=-3mm,
        font=\normalsize
    }
}

	\else
		
	\fi
	\tikzset{external/force remake}	
\fi

\newcommand{\Langold}{{\Lang}_{0}}
\newcommand{\Langnew}{{\Lang}_{1}}
\newcommand{\TTR}{{\sf TTR} }
\newcommand{\TTRi}{\mbox{$\mathsf{TTR}$} }
\newcommand{\WTR}{{\sf WTR}}
\newcommand{\WTRi}{\mbox{$\mathsf{WTR}$}}
\newcommand{\ts}{\mathsf{ts}}
\renewcommand{\imp}{\Rightarrow}
\newcommand{\trigg}{e_{\mathsf{s}}}

\title{Agent-time Epistemics and Coordination}
\author{Ido Ben-Zvi\\
  Department of Electrical Engineering, Technion\\
  \texttt{idobz@cs.technion.ac.il}
  \and 
  Yoram Moses\\
  Department of Electrical Engineering, Technion\\
  \texttt{moses@ee.technion.ac.il}  
  }
\begin{document}
\maketitle

\begin{abstract}
A minor change to the standard epistemic logical language,
replacing $K_{i}$ with $K_{\node{i,t}}$ where $t$ is a time instance, gives rise to a generalized 
and more expressive form of knowledge and common knowledge operators. We investigate the 
communication structures that are necessary for such generalized epistemic states to arise, 
and the inter-agent coordination tasks that require such knowledge. 
Previous work has 
established a relation between linear event ordering and nested knowledge,
and between simultaneous event occurrences and common knowledge. 
In the new, extended, formalism, 
epistemic necessity is decoupled from temporal necessity. 
Nested knowledge and event ordering are shown to be related even
when the nesting order does not match the temporal order of occurrence.
The generalized form of common knowledge does {\em not} correspond to simultaneity.
Rather, it corresponds to a notion of tight coordination, of which simultaneity is an instance. 
\end{abstract}

\section{Introduction}\label{sec:intro}
We have recently embarked on an in-depth inquiry concerning the relation between knowledge, 
coordination and communication in multi agent systems~\cite{BzMDISC2010,BzMLOFT2010,BenZvi2011}. 
This study uncovered new structural connections between the three in 
systems where agents have accurate clocks, and there are  (commonly known) bounds on 
the time it may take messages to be delivered, between any two neighboring agents. We call this the 
{\em synchronous} model. In such a setting, one often reasons about what agents know at particular time
points in the past or future, as well as, in particular, what they know about 
what other agents will know (or have known) at various other times, etc. 

The emphasis on information regarding the times at which facts are known 
has lead us to consider a formalization in which epistemic operators are indexed 
by a pair consisting of an agent {\em and a time}---thus, $K_{\node{i,t}}$ refers to what~$i$
knows at time~$t$---rather than the more traditional epistemic operator~$K_i$, in which knowledge 
is associated with an agent, and the formulas are evaluated with respect to 
a particular time. An agent-time pair $\node{i,t}$ is called a {\em node}, and we 
distinguish the new {\em node-based} (or {\em nb-}) language from the traditional {\em agent-based} one. 

In this paper we formulate an expressive language with nb-knowledge operators, 
and use it to extend and strictly strengthen 
the previously established relations between knowledge, 
coordination and communication. 
Our earlier inquiries reduced coordination tasks to states of knowledge, and then 
analyzed the communication requirements required to obtain these states of 
knowledge. Together these provide a crisp structural characterization of necessary and sufficient 
conditions for particular coordination tasks, in terms of the communication required. 

We specify coordination tasks in terms of orchestrating a pattern of responses 
in reaction to a spontaneous external input initiated by the environment. 
Such a spontaneous event is considered as the {\em trigger} for its responses. 
As shown in~\cite{BzMDISC2010}, ensuring a linearly ordered sequence of responses 
to such a triggering event requires a state of deeply nested knowledge, in which 
the last responder to know that the next-to-last responder 
knows, \ldots, that the agent performing the first response knows that the trigger 
event occurred.
In {\em asynchronous} systems without clocks, such nested knowledge can only be obtained via a message chain 
from the trigger passing through the responders in their order in the sequence. 
The situation in the synchronous setting, in the presence of timing information,  is much more delicate and interesting. In that case, such nested 
knowledge requires a particular pattern of communication called a  \emph{centipede}.
While nested knowledge is captured by a centipede, common knowledge corresponds 
to a more restrictive and instructive 
communication pattern called a \emph{broom}\footnote{A structure called {\em centibroom} 
in~\cite{BzMDISC2010} has since 
been renamed to be a broom.}). Using the connection between common knowledge and simultaneity, 
brooms are shown to underly natural forms of simultaneous coordination. 

The  node-based operators give rise to natural strict generalizations of 
nested and common knowledge. Of particular interest
is node-based common knowledge, which is represented by an epistemic operator $C_{A}$, in which $A$ is 
a set of agent-time pairs $\node{i,t}$. So $\,C_{\{\node{i,t},\node{j,t+3}\}}\vphi$ ~indicates, among
other things, that agent $i$ at time $t$ knows that agent $j$ at time $t+3$ will know that
$i$ at time $t$ knew that $\vphi$ holds. 
While nb-common knowledge strictly generalizes classical common knowledge, 
it shares most of the typical attributes associated 
with common knowledge. However, whereas (traditional) common knowledge is intimately related to 
simultaneity, nb-common knowledge {\em is not!}
We will show that, instead, nb-common knowledge precisely captures 
a form of coordination we call tightly-timed response~\mbox{(\TTR).}
Of course, simultaneity is a particular form of tightly-timed  coordination.

In general, we follow the pattern of investigation that we established in earlier papers:
Given an epistemic state (such as nb-nested or nb-common knowledge) we examine, on the one hand,
what form of communication among the agents is necessary for achieving such a state, and on the
other  we characterize the type of coordination tasks that require such a state.
As nb-knowledge operators generalize the standard agent-based ones, the results we obtain here
generalize and extend the ones of~\cite{BzMDISC2010,BenZvi2011}.
Relating knowledge to communication, we present two ``knowledge-gain'' theorems 
(formulated in the spirit of Chandy and Misra's result for systems without clocks \cite{ChM}).
Nb-nested and nb-common knowledge are shown to require more flexible variants of the centipede 
and the broom communication structures (respectively), in order to arise.  

Perhaps more instructive is the characterization of the coordinative tasks 
related to these epistemic states. As mentioned above, nested knowledge characterizes
agents that are engaged in responding in a sequential order 
to a triggering event. It turns out that Nb-nested knowledge is similarly necessary when
the ordering task predefines not only the sequence of responses, but also a bound
on the elapsed time between each response and its successor. We define this
task as the \emph{Weakly Timed Response} problem (\WTR). 
Interestingly, minute
differences in the coordinative task specification result in dramatic changes to the
required epistemic state. As for common knowledge, as hinted above it will be seen
that the node-based version relaxes the typical simultaneous responses requirement
into tightly-timed ones. 

The main contributions of this paper are:
\begin{itemize}
\item Node-based epistemic operators are defined. Under a natural semantic definition, 
both nb-knowledge and nb-common knowledge are S5 operators, while 
nb-common knowledge satisfies fixed-point and induction properties analogous to those of 
standard common knowledge.
\item The theory of coordination and its relation to epistemic logic are extended, 
significantly generalizing previous results. Nb-semantics helps to decouple
epistemic and temporal necessity. Namely, there are cases in which knowledge about 
(the guarantee of) someone's knowledge at a future time, typically based on known communication,
is required in order to perform an earlier action. 
\item The new epistemic operators are used as a formal tool for the study of 
generalized forms of coordination and the communication structures they 
require. On the one hand, natural coordination tasks that require nb-nested 
and nb-common knowledge are identified. On the other hand, the communication 
structures necessary for attaining these nb-epistemic states are established. 
Combining the two types of results yields new structural connection between 
communication and coordination tasks. 
%
%
\item The well-known strong connection between common knowledge and simultaneity
established in \cite{HM1,FHMV,FHMV96}  is shown to be a particular instance of a more 
general phenomenon: We prove that tightly-timed synchronization is very closely related 
to node-based common knowledge. The form of tight synchronization corresponding 
to agent-based common knowledge is precisely simultaneous coordination. 
\end{itemize}

This paper is structured as follows. In the next section we define the syntax and semantics of two languages: 
One is a traditional agent-based logic of knowledge, and the other a node-based logic.
An embedding of agent-based formulas in the nb-language is established, and 
basic properties of nb-knowledge and nb-ck are discussed.
Section~\ref{sec:model} presents the synchronous model within which we will study knowledge and coordination.
In Section~\ref{sec:background} a review of the notions underlying the earlier analysis, and its main results,
is presented. 
Our new analysis is presented in Sections~\ref{sec:WTR} and~\ref{sec:TTR}. 
Section~\ref{sec:WTR} defines ``uneven'' centipedes and relates them to nested nb-knowledge. It then defines 
a coordination task called {\em weakly-timed response} and shows that it is captured by nested nb-knowledge.
Analogously, 
Section~\ref{sec:TTR} defines uneven brooms and relates them to nested nb-common knowledge. 
It then defines {\em tightly-timed response} and shows that it is captured by nested nb-ck.
Finally, a short discussion and concluding remarks and presented in Section~\ref{sec:conc}.
Proofs of the claims in Sections~\ref{sec:WTR} and~\ref{sec:TTR} are presented in the Appendix.

\section{Agent and Node-based Semantics}\label{sec:nb-semantics}
We consider both the standard, agent-based, epistemic language and its extension
into the node-based variant within the \emph{interpreted systems} framework
of~\cite{FHMV}. In this framework, the multi agent system is viewed 
as consisting of a set $\Proc=\{1,\ldots,n\}$  of
agents, connected by a communication network which serves as the exclusive
means by which the agents interact with each other. 

We assume that, at any given point
in time, each agent in the system is in some {\em local state}. A {\em global
state\/} is just a tuple $g=\langle\ell_e, \ell_1, \ldots, \ell_n\rangle$
consisting of local states of the agents, together with the state~$\ell_e$ of
the {\em environment}. The environment's state accounts for everything that is
relevant to the system that is not contained in the state of the agents.
A {\em run\/} $r$ is a function from time to global states. Intuitively, a run is a
complete description of what happens over time in one possible execution of the
system.  We use
$r_i(t)$ to denote agent $i$'s local state $\ell_i$ at time $t$ in run $r$, for 
$i = 1,\ldots, n$.  For simplicity, time here
is taken to range over the natural numbers rather than the reals (so that time
is viewed as discrete, rather than dense or continuous). {\em Round\/} $t$ in
run $r$ occurs between time $t-1$ and $t$.
A \emph{system} $\Rrep$ is an exhaustive set of all possible runs, given the
agents' protocol and the {\em context}, where the latter determines 
underlying characteristics of the environment as a whole.

To reason about the knowledge states of agents, a simple logical language is introduced.
Since we are focusing on coordination tasks in which actions are triggered by 
spontaneous events, the only primitive propositions we consider are ones that state 
that an event has occurred. 
The standard,  agent-based, variant of this language is $\Langold$. It uses the following grammar, 
with the usual abbreviations for $\vee$ and $\imp$, and with $e,i$ and $G$
used as terms for an event, an agent and a group of agents, respectively.
\vspace{-4pt}
\[\vphi \qquad ::= \qquad \occurred{e} ~|~ \vphi \wedge \vphi ~|~\neg \vphi ~|~
K_{i}\vphi ~|~ E_{G}\vphi ~|~ C_{G}\vphi\]

The semantics is as follows. Propositional connectives, omitted from the list, are given
their usual semantics.

\begin{definition}[$\Langold$ semantics]	
	The truth of a formula $\vphi\in \Langold$ is defined with respect
	to a triple $(\Rrep,r,t)$.
	\begin{itemize}
		\item $(\Rrep,r,t)\sat\occurred{e}$ ~iff ~event~$e$ has occurred in $r$ by time $t$. 
		\item $(\Rrep,r,t)\sat K_i\varphi$ ~iff ~$(\Rrep,r',t)\sat \varphi$
		  for all $r\in\Rrep$ s.t.\ $r_i(t)=r'_i(t)$.
		\item $(\Rrep,r,t)\sat E_G\varphi$ ~iff ~$(\Rrep,r,t)\sat K_i\varphi$
		  for every $i \in G$.
		\item $(\Rrep,r,t)\sat C_G\varphi$ ~iff ~$(\Rrep,r,t)\sat
		  (E_G)^k\varphi$ for every $k\ge 1$.
	\end{itemize}
\end{definition}

Note that while non-epistemic formulas require only a run $r$ and  time $t$ in order
to be evaluated, the epistemic operators also require the complete
system of runs $\Rrep$ (which provides the analogue of the set of states, or {\em possible worlds} in standard modal logic). 
The second clause, giving semantics for the knowledge operator~$K_i$, has built-in 
an assumption that time is common knowledge, since knowledge at time~$t$ in~$r$ depends
only on truth in other runs, that are indistinguishable  for agent~$i$ from~$r$ at {\em the same time~$t$}. 
This is appropriate for our intended analysis, since in our model (to be presented in Section~\ref{sec:model}) 
time is assumed to be common knowledge. 
We used $\Langold$ to formalize the analysis in~\cite{BzMDISC2010}.

We next define the node-based language $\Langnew$. 
It will be convenient to define the set  $\CV=\{\node{i,t}: i\in \Proc, t\textrm{ is a time}\}$
of all possible nodes.
In the grammar we now use $e,\alpha$ and $A$
 as terms for an event, an agent-time node and a group of such nodes, respectively.
\vspace{-4pt}
\[\vphi \qquad ::= \qquad \tocc{t}{e}  ~|~ \vphi \wedge \vphi ~|~ \neg \vphi ~|~
K_{\alpha}\vphi~|~E_{A}\vphi~|~C_{A}\vphi\]

Notice that all statements in this language are ``time-stamped:'' they refer to explicit times at 
which the stated facts hold. As a result, they are actually time-invariant, and state facts about 
the run, rather than facts whose truth depends on the time of evaluation. 
Therefore, semantics for formulas of $\Langnew$ are given with respect to a system~$\Rrep$ and
a run $r\in\Rrep$. 
The semantics is as follows (again. ommiting propositional clauses). 

\begin{definition}[$\Langnew$ semantics]	
	The truth of a formula $\vphi\in \Langnew$ is defined with respect
	to a pair $(\Rrep,r)$.
	\begin{itemize}
		\item $(\Rrep,r)\sat\tocc{t}{e}$ iff event~$e$ has occurred in $r$ by time $t$ 
		\item $(\Rrep,r)\sat K_\node{i,t}\varphi$ iff~$(\Rrep,r')\sat \varphi$
		  for all runs~$r' \textrm{ s.t. }r_i(t)=r'_i(t)$.
		\item $(\Rrep,r)\sat E_A\varphi$ iff~$(\Rrep,r)\sat K_\alpha\varphi$
		  for every $\alpha \in A$.
		\item $(\Rrep,r)\sat C_A\varphi$ iff~$(\Rrep,r)\sat
		  (E_A)^k\varphi$ for every $k\ge 1$.
	\end{itemize}
\end{definition}

In principle, the node-based semantics, as proposed here, can be seen
as a simplified version of a real-time temporal logic with an explicit clock variable~\cite{Alur92},  
and more generally of a  hybrid logic~\cite{Areces05}. We are not aware of instances in which 
such logics have been combined with epistemic operators.
Nevertheless, in this paper we aim to utilize the formalism, rather than explore it.
Hence, issues of expressibility, completeness
and tractability are left unattended, to be explored 
at  a  future date.   

We now demonstrate that, in a precise sense, the traditional language $\Langold$ can be embedded 
in the language $\Langnew$, in a meaning-preserving manner. 
We do this by way of defining a {\em ``timestamping''} operation $\ts$ transforming a formula $\vphi\in\Langold$ 
and a time~$t$ to a formula $\ts(\vphi,t)=\vphi^t\in\Langnew$ that is timestamped by~$t$.  We then prove 
the following lemma showing that the timestamping is sound.

\begin{lemma}
\label{lem:timestamping}
	There exists a function $\ts: \Langold \times Time \to \Langnew$
	such that for every $\vphi\in\Langold$ , time $t$, and $\vphi^{t}=\ts(\vphi,t)$:
	$$(\Rrep,r,t)\sat \vphi \qquad \textrm{iff} \qquad (\Rrep,r)\sat \vphi^{t}.$$
\end{lemma}

Lemma~\ref{lem:timestamping} shows that $\Langnew$ is at least as expressive as $\Langold$. 
Yet $\Langnew$ is not equivalent in expressive power to $\Langold$, since 
it allows for multiple temporal reference points where  epistemic operators
are involved. For example, the formula $K_{\node{i,t}}K_{\node{j,t'}}\vphi$ cannot
be translated into an equivalent $\Langold$ formula, since $\Langold$
does not allow for the temporal reference point
to be shifted  when switching from the outer knowledge operator to the inner one.

Compare a typical agent-based nested knowledge formula, such as $\psi=
K_{i}K_{j}\vphi$, with a node based counterpart such as $\psi' =
K_{\node{i,t}}K_{\node{j,t+3}}\vphi$. In~$\psi$, gaining knowledge about the
epistemic state of other agents means that the knowledge of the referred other
agent has already been gained. 
By contrast, in~$\psi'$ agent $i$ has
epistemic certainty concerning the knowledge state of another agent $j$, 
at a particular time in the future. 
To sum up, the node-based formalism allows us to differentiate between epistemic necessity, and
temporal tense. Formally trivial, this decoupling is quite elusive as we tend to
mix one with the other. The rest of this paper follows up on the
implications of this point.

It is well-known that standard agent-based common knowledge is closely related to simultaneity
\cite{HM1,FHMV,FHMV96}.
Indeed, both $C_G\vphi \imp E_GC_{G}\vphi$ and  $K_iC_G\vphi\imp C_G\vphi$ are valid formulas. 
(Recall that a formula $\psi\in\Langold$ is {\em valid} if $(\Rrep,r,t)\sat\psi$ for all choices of $\Rrep$, run $r\in\Rrep$ 
and time~$t$.) 
Thus, the first instant at which $C_G\vphi$ holds must involve a simultaneous change in the local states
of all members of~$G$. 
In contrast, simultaneity is {\em not} an intrinsic property of node-based common knowledge.
As an example, consider the node set 
$A=\{\node{i,t},\node{j,t+10}\}$. Although
we still have that both $K_{\node{i,t}}C_{A}\vphi \imp C_{A}\vphi$ and 
$C_A\vphi \imp K_{\node{j,t+10}}C_{A}\vphi$ are valid formulas, 
if the current time is $t$ and $i$ knows that $C_{A}\vphi$
holds, this does not mean that $j$ currently knows this too. The bond with simultaneity 
has been broken. 
As we shall see in Section~\ref{sec:TTR}, however, 
a notion of tight temporal coordination that generalizes simultaneity {\em is} intrinsic 
to node-based common knowledge. 

In the spirit of the treatment in~\cite{FHMV}, for a formula $\psi\in\Langnew$ we write $\Rrep\sat\psi$ 
and say that $\psi$ is {\em valid in} (the system) $\Rrep$ if $(\Rrep,r)\sat\psi$ for all $r\in\Rrep$. 
Node-based common knowledge manifests many
of the logical properties shown by the standard notion of common knowledge. 
Proof of the 
following lemma, and of all new lemmas in this paper, can be found in the Appendix.

\begin{lemma}\label{lem:prop-nodeCK}\mbox{}
	\begin{itemize}
		\item Both $K_{\node{i,t}}$ and $C_A$ are $S5$ modalities.
		\item $C_{A}\vphi \imp E_{A}(\vphi \wedge C_{A}\vphi)$ is valid.
		\item If ~$\Rrep\sat \vphi \imp E_{A}(\vphi \wedge \psi)$ ~then ~$\Rrep\sat \vphi \imp C_{A}\psi$
	\end{itemize}
\end{lemma}
The second clause of Lemma~\ref{lem:prop-nodeCK} corresponds to the so-called ``fixed-point'' 
axiom of common knowledge, while the third clause corresponds to the ``induction (inference) rule''
\cite{FHMV}.

\section{The Synchronous Model}\label{sec:model}
As mentioned, we focus on synchronous systems where 
the clocks of the individual agents are all synchronized, and there
are  commonly known bounds on message delivery times. 
Generally, in order to define a system of runs that conforms to the required 
setting of interest, we define the system of runs $\Rrep$ as a function 
$\Rrep=\Rrep(P,\gamma)$ of the
protocol $P$ followed by the agents, and the underlying context of use $\gamma$.

We identify a {\em protocol\/} for an agent~$i$ with a function from local
states of~$i$ to nonempty sets of actions. (In this paper we assume a {\em
deterministic\/}  protocol, in which a local state is mapped to a singleton set
of actions. Such a protocol essentially maps local states to actions.) A
\emph{joint protocol} is just a sequence of protocols $P=(P_1,\ldots,P_n)$, one
for each agent.

In this paper we will assume a specified protocol for the agents, namely the
$\fip$, or \emph{full information protocol}. In this protocol, every agent sends
out its complete history to each of its neighbors, on every round. In order to be able 
to do that, the agent must be able to recall its own history. We therefore also
assume this capability, called \emph{perfect recall}, for the agents. Note that using
a pre-specified protocol is not in general necessary, and is done
for deductive purposes only. Our findings can be
applied to all protocols under slight modifications.

In order for a well defined system of runs $\Rrep$ to emerge, the context
$\gamma$ needs to be rigorously defined as well. 
We also assume a specific context $\gammax$, within which the agents are operating.
Most notably, $\gammax$ specifies that (a) agents share a universal notion of time, 
(b) the communication network has upper bounds on delivery times, and (c) all nondeterminism
is deferred to the environment agent.

A word regarding nondeterminism, that plays a crucial role in the analysis of
knowledge gain. Formal analysis allows us to escape the more difficult questions
associated with this concept. For us, intuitively, spontaneous events are
ones that cannot be foretold by the agents in the system. These could stand for
a power shortage, but also for some user input that is communicated to an agent via
a console. Formally, since all of the agents are following a deterministic
protocol, nondeterminism is only introduced into the system by the environment.

For brevity, we must omit the formal details\footnote{The interested reader can find 
complete accounts of the context we assume in~\cite{BzMDISC2010,BenZvi2011}},
and make do with describing the properties of the outcome system $\Rrep=\Rrep(\fip,\gammax)$:

\begin{itemize}
	\item Global clock and global network - The current time is always common knowledge.
	The network, which can be encoded as a weighted graph, is also common knowledge.
	For each $(i,j)$ connected by a communication channel, $\bij$, the weight on the corresponding
	network edge, denotes the maximal transmission times for messages sent along
	this channel. 
	\item Events - There are four kinds of events: message send and receive events,
	internal calculations, and external inputs. Events occur at a single agent, within 
	a single round. All are self explanatory except for the last type. External input events
	occur when a signal is received by an agent from ``outside'' the system. This could
	be user input, fate, and also - less dramatically - it could be used to 
	signify the initial values for internal variables at the beginning of a run. 
	\item Environment protocol - The environment is responsible for the
	occurrence of two of the event types: message deliveries and external inputs.
	(a) message deliveries - the protocol dictates that message deliveries will occur only for messages
	that have been sent and are still en route. Apart from that, the environment 
	nondeterministically chooses when to deliver sent messages, subject only
	to the constraint that for every communication channel $(i,j)$, transmission
	on the channel does not take longer than $\bij$ rounds. (b) external inputs - 
	in each round the environment nondeterministically chooses a (possibly empty) set of agents
	at which external inputs will occur, and the kind of events that will occur there. 
	The choice is entirely unconstrained by previous or
	simultaneous occurrences. 	
\end{itemize}

\section{Previous Findings}\label{sec:background}
Our approach is based on the findings of Lamport~\cite{LamClocks}
and of Chandy and Misra~\cite{ChM}. In his seminal analysis,
Lamport defines \emph{potential causality}, a formalization
of  message chains. Two events are related by potential causality when the
coordinates marking their occurrences (the time and agent at which they
occur) are related by an unbroken sequence of  messages. We
rephrase the relation as one holding between pairs of agent-time nodes.

\begin{definition}[Potential causality~\cite{LamClocks}]
	\label{def:lam}
	Fix $r\in\Rrep$. The {\em potential causality} relation $\spc$ over nodes
	of~$r$ is the smallest relation satisfying the following three conditions:
	\begin{enumerate}
		\item If ~$t\leq t'$  then $\node{i,t}\spc \node{i,t'}$;
		\item If some message is sent at $\node{i,t}$ and received at $\node{j,t'}$
		then $\node{i,t}\spc \node{j,t'}$; and
		\item If ~$\node{i,t}\spc \node{h,t''}$ and ~$\node{h,t''}\spc \node{j,t'}$,
		then ~$\node{i,t}\spc \node{j,t'}$.
	\end{enumerate}
\end{definition}   

Lamport used the relation as a basis for his \emph{logical clocks}
mechanism that allows distributed protocol designers in asynchronous 
contexts to temporally order events despite the lack of synchronization.
Chandy and Misra later gave an epistemic analysis, showing that 
for agent $j$ at time $t'$ to know of an occurrence at agent $i$'s at time
$t$, it must be that $\node{i,t}\spc\node{j,t'}$ holds.

In~\cite{BzMDISC2010,BzMLOFT2010,BenZvi2011} 
we  applied similar methodology to analyze knowledge
gain and temporal ordering of events in \emph{synchronous} systems, like the one described here. 
Note that temporally ordering events in a system with global clock is quite
easy if exact timing is prearranged as a part of the protocol: 

\emph{Charlie will deposit the money at 3pm, you will sign the contract at 4, 
and I will deliver the merchandise at 5}.

Coordination becomes more challenging once the time of occurrence
of the triggering, initial, action is nondeterministic:

\emph{Charlie will deposit the money, err... whenever, then you will sign the contract, 
and only then will I  deliver the merchandise}. 

To approach the issue of event ordering we define the
\emph{Ordered Response} coordination challenge. Such
problems are defined based on a \emph{triggering event} $\trigg$, which
is an event of type external input, and a set of \emph{responses} 
$\rsp{\alpha_1},\ldots,\rsp{\alpha_k}$.
Each response $\rsp{\alpha_{h}}$ is a pair $\angles{i_{h},a_{h}}$ indicating an action $a_{h}$ that agent
$i_{h}$ is required to carry out. When a  response $\rsp{\alpha_{h}}$ gets carried out 
in a specific run, we  denote with $\alpha_{h}=\node{i_{h},t_{h}}$ the specific agent-time 
node at which action $a_{h}$ occurs.

\begin{definition}[Ordered Response~\cite{BzMDISC2010}] \label{def:OR}
	Let $\trigg$ be an external input nondeterministic event.
	A protocol~$P$ solves the instance
	$\ORi=\angles{\rsp{\trigg},\rsp{\alpha_1},\ldots,\rsp{\alpha_k}}$ of the {\em Ordered Response}
	problem if it guarantees that
	\begin{enumerate}
		\item every response $\rsp{\alpha_h}$, for $h=1,\ldots,k$, occurs in a run
		iff the trigger event $\trigg$ occurs in that run.
		\item for  $h=1,\ldots,k-1$, response $\rsp{\alpha_{h}}$ occurs no later
		than response $\rsp{\alpha_{h+1}}$ (i.e. $t_{h}\leq t_{h+1}$).
	\end{enumerate}
\end{definition}

We can show that solving the challenge for me, you and Charlie 
requires that you will know that Charlie has deposited 
before you sign the contract, and that I will know that you know that Charlie
has made the deposit, before I deliver the merchandise. Theorem~\ref{thm:ORnestedK}
below proves this link between coordination and knowledge. 
Like all of the results quoted from previous works in this section, it is expressed
using the language $\Langold$.

\begin{theorem}[\cite{BenZvi2011}]
\label{thm:ORnestedK}
	Let \ORi=$\angles{\trigg,\rsp{\alpha_1},\ldots,\rsp{\alpha_k}}$ be an instance of \OR,
	and assume that \ORi is solved in the system $\Rrep$.
	Let $r\in\Rrep$ be a run in which~$\trigg$ occurs, let $1\le h\le k$,
	and let $t_h$ be the
	time at which $i_h$ performs action~$a_h$ in~$r$. Then
	\[(\Rrep,r,t_h)\sat~K_{i_h}K_{i_{h-1}}\cdots K_{i_1}\occurred{\trigg}.\]
\end{theorem}

Given that nested knowledge can be shown to be a necessary prerequisite
for ordering actions, what patterns of communication will provide such knowledge
gain? One way to ensure event ordering is to insist on a message chain, linking
Charlie to you and then to me. In fact, as Chandy and Misra show, it is the \emph{only}
way to ensure this, in an asynchronous system. But the  synchronous setting 
allows for more flexibility in ensuring event ordering.  For example, we could have: 

\emph{Charlie sends messages to both you and me, alerting us of the deposit.
Given that transmission times are bounded from above, upon getting a message from
Charlie I  calculate how long before the message sent to you is guaranteed to arrive, and
deliver the merchandise only then}. 

The extra flexibility, when compared to  asynchronous systems, is based on the 
availability of guarantees on message delivery times.
We formalize these guarantees in the following way:
 
\begin{definition}[Bound guarantee~\cite{BzMDISC2010}]
	\label{def:bound}
	Fix $r\in\Rrep$. The {\em bound guarantee} relation $\tspc$ over nodes
	of~$r$ is the smallest relation satisfying the following three conditions:
	\begin{enumerate}
		\item If ~$t\leq t'$  then $\node{i,t}\tspc \node{i,t'}$;
		\item If the network contains a channel $(i,j)$ with weight $\bij$
		~then $\node{i,t}\tspc \node{j,t+\bij}$; and
		\item If ~$\node{i,t}\tspc \node{h,t''}$ and ~$\node{h,t''}\tspc \node{j,t'}$,
		then ~$\node{i,t}\tspc \node{j,t'}$.
	\end{enumerate}
\end{definition}   

With most of the building blocks in place, we are almost ready to describe the Knowledge Gain Theorem,
the technical result at the core of our~\cite{BzMDISC2010} paper. The theorem characterizes
the communication pattern that is necessary for nested knowledge gain in 
synchronous systems. Intuitively, it shows that message chains still play an
important part in information flow, but the synchronous equivalent of 
a message chain is much more flexible - since here agents can also 
use bound guarantees in order to ensure that a message arrives at its destination.
The resultant message chain abstraction, called the \emph{centipede}, is defined below.

\begin{figure}[!h]
\centering
\fbox{
\newcommand{\tikzfname}{tikz_centipede}
\tikzsetnextfilename{Figures/\tikzfname}
\resizebox{0.6\linewidth}{!}{ \begin{tikzpicture}[font=\huge]
        \node[agent_time=\nodeC, label=left:{$\alpha_{0}$}]  (i0) at (0,0) {};
        	\coordinate (ct) at (4,1) {};
        	\coordinate (c1) at (6,2.4) {};
        	\coordinate (c2) at (6,1.7) {};
        	\coordinate (c3) at (6,1) {};
        	\coordinate (ck) at (6,-0.5) {};
        
        \path (i0) -- (ct) node[pos=0.9,agent_time=\nodeC,  label=above:{$\theta$}] (t)  {};   
        \path (t) -- (c1) node[pos=1.2,agent_time=\nodeC, label=right:{$\alpha_{1}$}] (i1)  {};   
        \path (t) -- (c2) node[pos=1,agent_time=\nodeC, label=right:{$\alpha_{2}$}] (i2)  {};   
        \path (t) -- (c3) node[pos=1.6,agent_time=\nodeC, label=right:{$\alpha_{3}$}] (i3)  {};   
        \path (t) -- (ck) node[pos=1.3,agent_time=\nodeC, label=right:{$\alpha_{k}$}] (ik)  {};

        \draw[loosely dotted, ultra thick, shorten <=6pt, shorten >=6pt] (c3) to (ck);

          \draw[syncausality] (i0) to (t);

           \draw[guarantee] (t) to (i1);
           \draw[guarantee] (t) to (i2);
           \draw[guarantee] (t) to (i3);
           \draw[guarantee] (t) to (ik);
 \end{tikzpicture}}
}
\caption{A centipede}
\label{fig:centipede}
\end{figure}
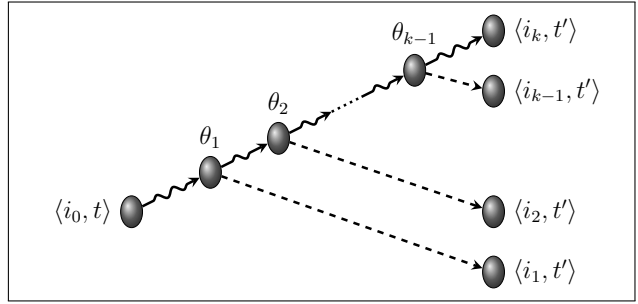

\begin{definition}[Centipede~{\cite{BzMDISC2010}}]
\label{def:cpede}
	Let $r\in\Rrep$, let $i_h\in\Proc$ for $0\le h\le k$ and let $t\le t'$.
	A {\em centipede} for $\langle i_0,\ldots,i_k\rangle$ in the interval
	$[t,t']$ in $r$ is a
	sequence $\langle\theta_0,\theta_1,\ldots,\theta_k\rangle$ of nodes 
	such that $\theta_0=\node{i_0,t}$, $\theta_k=\node{i_k,t'}$,
	~$\theta_0\spc\theta_1\spc\cdots\spc\theta_k$, 
	and ~\mbox{$\theta_h\tspc (i_h,t')$} ~holds for $h=1,\ldots, k-1$.
\end{definition}

A centipede for $\langle i_0,\ldots,i_k\rangle$  is
depicted in Figure~\ref{fig:centipede}. It shows a
message chain connecting $(i_0,t)$ and $(i_k,t')$, and along this chain a
sequence of ``route splitting" nodes $\theta_1,\theta_2,$ etc. such that each
$\theta_h$ can guarantee the arrival of a message to $i_h$ by time $t'$. Such a
message can serve to inform $i_h$ of the occurrence of a trigger event at
$\node{i_{0},t}$, and as the set of previously made guarantees gets shuffled on
to the next splitting node, the last agent $i_k$ can be confident that by time $t'$
 all previous agents are already informed of the occurrence.

\begin{theorem}[Knowledge Gain Theorem~{\cite{BzMDISC2010}}]
\label{thm:NKgain}
    Let~$r\in\Rrep$.
    Assume that $e$ is an external event occurring at $\node{i_0,t}$ in $r$.\\
    If $(\Rrep,r,t')\sat K_{i_k}K_{i_{k-1}}\!\!\cdots
    K_{i_{0}}\occurred{e}$, 
    then there is a centipede for $\langle i_0,\ldots,i_k\rangle$ in
    the interval $[t,t']$ in $r$.
\end{theorem}

Theorem~\ref{thm:ORnestedK} states that solving the ordering problem
requires nested knowledge of the occurrence of the trigger event to have
been gained. Theorem~\ref{thm:NKgain} then shows that such knowledge gain
can only take place if the agents are related by the centipede communication
pattern. 

Although stated in terms of our system $\Rrep$, where agents
are assumed to be following $\fip$, under some further generalization
both theorems can be made to apply for all systems
based on the synchronous context $\gammax$, regardless of the specific protocol.\footnote{
In order to generalize the theorems in this way we need to extend potential
causality into a relation called \emph{syncausality}. See~\cite{BzMDISC2010} for more.}

The ordering problem, nested knowledge and the centipede define a ``vertical stack'',
going from coordination, to knowledge, to communication and centered about nested knowledge
gain. We now examine another such vertical stack, this time defined based on common knowledge.
Common knowledge has been associated with simultaneous action already in~\cite{FHMV,HM1}. We 
touch upon this relation  by defining the \emph{Simultaneous Response} problem and then using
Theorem~\ref{thm:SRck} below.

\begin{definition}[Simultaneous Response~{\cite{BzMDISC2010}}]
	Let $\trigg$ be an external input nondeterministic event.
	A protocol~$P$ solves the instance
	$\SRi=\angles{\trigg,\rsp{\alpha_1},\ldots,\rsp{\alpha_k}}$ of the {\em Simultaneous Response} 
	problem if it guarantees that
	\begin{enumerate}
		\item every response $\rsp{\alpha_h}$, for $h=1,\ldots,k$, occurs in a run
		iff the trigger event $\trigg$ occurs in that run.
		\item all of the responses $\rsp{\alpha_{1}},\ldots,\rsp{\alpha_{k}}$ are performed simultaneously
		(i.e. $t_{1}=t_{2}=\cdots =t_{k}$).
	\end{enumerate}
\end{definition}
 
\begin{theorem}[{\cite{BzMDISC2010}}]\label{thm:SRck}
	Let $\SRi=\angles{\trigg,\rsp{\alpha_1},\ldots,\rsp{\alpha_k}}$,
	and assume that $\SRi$ is solved in~$\Rrep$.
	Moreover, let $G=\{i_1,\ldots,i_k\}$ be the set of processes appearing
	in the response set of~$\SRi$.
	Finally, let $r\in\Rrep$ be a run in which~$\trigg$ occurs,
	and let $t$ be the time at which the response actions are performed
	in~$r$. Then ~~$(\Rrep,r,t)\sat~C_G \occurred{\trigg}.$
\end{theorem}

Relating common knowledge to a necessary communication pattern is more
difficult, especially if you consider the formally-convenient classification of
common knowledge as an infinite conjunction of nested mutual knowledge
operators. Intuitively, borrowing the centipede from the nested knowledge
results quotes above, this would mean that an infinite centipede is required in order for
common knowledge to arise. However, it turns out that the necessary
communication pattern is quite finite, and more in line with a fixed point view:
In order for the group of agents $G$ to gain common knowledge that $\trigg$ has
occurred, there must exist a single agent-time node $\theta$ that is message
chain related to the site of occurrence, and that can guarantee that forwarded
messages with information regarding $\trigg$ to all members of $G$ will arrive by
$t'$. We call this pattern the \emph{broom}. The following definition, and Figure~\ref{fig:broom}
below, should make it apparent why.

\begin{figure}[!h]
\centering
\fbox{
\newcommand{\tikzfname}{tikz_broom}
\tikzsetnextfilename{Figures/\tikzfname}
\resizebox{0.6\linewidth}{!}{ \begin{tikzpicture}[font=\huge]
        \node[agent_time=\nodeC, label=left:{$\alpha_{0}$}]  (i0) at (0,0) {};
        	\coordinate (ct) at (4,1) {};
        	\coordinate (c1) at (6,2.4) {};
        	\coordinate (c2) at (6,1.7) {};
        	\coordinate (c3) at (6,1) {};
        	\coordinate (ck) at (6,-0.5) {};
        
        \path (i0) -- (ct) node[pos=0.9,agent_time=\nodeC,  label=above:{$\theta$}] (t)  {};   
        \path (t) -- (c1) node[pos=1.2,agent_time=\nodeC, label=right:{$\alpha_{1}$}] (i1)  {};   
        \path (t) -- (c2) node[pos=1,agent_time=\nodeC, label=right:{$\alpha_{2}$}] (i2)  {};   
        \path (t) -- (c3) node[pos=1.6,agent_time=\nodeC, label=right:{$\alpha_{3}$}] (i3)  {};   
        \path (t) -- (ck) node[pos=1.3,agent_time=\nodeC, label=right:{$\alpha_{k}$}] (ik)  {};

        \draw[loosely dotted, ultra thick, shorten <=6pt, shorten >=6pt] (c3) to (ck);

          \draw[syncausality] (i0) to (t);

           \draw[guarantee] (t) to (i1);
           \draw[guarantee] (t) to (i2);
           \draw[guarantee] (t) to (i3);
           \draw[guarantee] (t) to (ik);
 \end{tikzpicture}}
}
\caption{A broom}
\label{fig:broom}
\end{figure}

\begin{definition}[Broom~{\cite{BzMDISC2010}}]
	\label{def:broom}
	Let $t\le t'$	and $G\subseteq\Proc$.
	Node $\theta$ is a {\em broom} for  $\angles{i_0,G}$ in the interval $[t,t']$ in $r$ if
	~$\node{i_0,t}\spc\theta$ ~and ~$\theta\tspc \node{i_h,t'}$ holds for
        all ~$i_h\in G$. 
\end{definition}

Theorem~\ref{thm:CKgain} states the necessity connection between common knowledge gain 
and the broom structure. As before, when we were dealing with nested knowledge,
Theorems~\ref{thm:SRck} and~\ref{thm:CKgain} can be made to hold for all synchronous ($\gammax$ based)
systems, regardless of protocol.
 
\begin{theorem}[Common Knowledge Gain~{\cite{BzMDISC2010}}]
	\label{thm:CKgain}
	Let~ $G\subseteq\Proc$, and let $r\in\Rrep$. Assume
	that   $e$ is an external input event at $\node{i_0,t}$ in~$r$. 
	If $(\Rrep,r,t')\sat C_G(\occurred{e})$,
	then there is a  broom $\hat \theta$ for $\angles{ i_0,G}$ in
	interval $[t,t']$ in $r$.
\end{theorem}

\section{Nested Knowledge and Weak Bounds}\label{sec:WTR}
We now re-approach the so-called ``vertical stack'' centered about nested knowledge.
This time we replace standard nested formulas such as $K_{i}K_{j}K_{k}\vphi$
with nb-formulations such as $K_{\node{i,t}}K_{\node{j,t-4}}K_{\node{k,t+8}}\vphi$.

What communication pattern is required in order to attain such
nested knowledge? The following generalization of the
centipede  echos  the above mentioned break between
temporal precedence and necessity. 

\begin{definition}[Uneven centipede]
\label{def:cpede}
	Let $r\in\Rrep$, let $A=\angles{\alpha_{0},\alpha_{1},\ldots,\alpha_{k}}$ be a sequence of nodes.
	An  {\em (uneven) centipede} for $A$ in $r$ is a
	sequence $\langle\theta_0,\theta_1,\ldots,\theta_k\rangle$ of nodes 
	such that $\theta_0=\alpha_{0}$, $\theta_k=\alpha_{k}$,
	~$\theta_0\spc\theta_1\spc\cdots\spc\theta_k$, 
	and ~\mbox{$\theta_h\tspc \alpha_{h}$} ~holds for $h=1,\ldots, k-1$.
\end{definition}

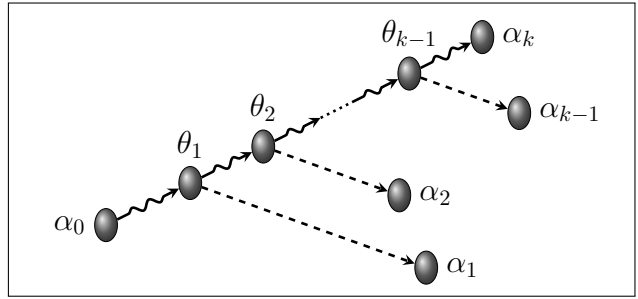
\begin{figure}[!h]
\centering
\fbox{
\newcommand{\tikzfname}{tikz_nbcentipede}
\tikzsetnextfilename{Figures/\tikzfname}
\resizebox{0.6\linewidth}{!}{ \begin{tikzpicture}[font=\huge]
        \node[agent_time=\nodeC, label=left:{$\alpha_{0}$}]  (i0) at (0,0) {};
        	\coordinate (ct) at (4,1) {};
        	\coordinate (c1) at (6,2.4) {};
        	\coordinate (c2) at (6,1.7) {};
        	\coordinate (c3) at (6,1) {};
        	\coordinate (ck) at (6,-0.5) {};
        
        \path (i0) -- (ct) node[pos=0.9,agent_time=\nodeC,  label=above:{$\theta$}] (t)  {};   
        \path (t) -- (c1) node[pos=1.2,agent_time=\nodeC, label=right:{$\alpha_{1}$}] (i1)  {};   
        \path (t) -- (c2) node[pos=1,agent_time=\nodeC, label=right:{$\alpha_{2}$}] (i2)  {};   
        \path (t) -- (c3) node[pos=1.6,agent_time=\nodeC, label=right:{$\alpha_{3}$}] (i3)  {};   
        \path (t) -- (ck) node[pos=1.3,agent_time=\nodeC, label=right:{$\alpha_{k}$}] (ik)  {};

        \draw[loosely dotted, ultra thick, shorten <=6pt, shorten >=6pt] (c3) to (ck);

          \draw[syncausality] (i0) to (t);

           \draw[guarantee] (t) to (i1);
           \draw[guarantee] (t) to (i2);
           \draw[guarantee] (t) to (i3);
           \draw[guarantee] (t) to (ik);
 \end{tikzpicture}}
}
\caption{A uneven centipede}
\label{fig:nb-centipede}
\end{figure}

Figure~\ref{fig:nb-centipede} shows such an uneven centipede. 
It is termed {\em uneven} because the ``legs'' of the centipede end at nodes 
at a variety of different times, whereas in the traditional centipede all legs ended 
at nodes of time~$t_k$. 
More interestingly, note that the 
node $\alpha_{2}$ temporally precedes $\alpha_{1}$ in Fig.~\ref{fig:nb-centipede}.   
Intuitively though, this does not seem to concur with the epistemic status of the two nodes, 
because information 
about the occurrence of the trigger event at $\alpha_{0}$
flows through $\theta_{1}$ and $\theta_{2}$ to $\alpha_{2}$, along with $\theta_{1}$'s
guarantee that by $t_{1}$ agent $i_{1}$  will have received word of the occurrence as well.
Thus, $K_{\alpha_2}K_{\alpha_1}\tocc{t_0}{\trigg}$ is attained by information flowing 
from a witness for $\alpha_1$ to the node~$\alpha_2$. Thus, in a precise sense, 
while $\alpha_1$ occurs temporally later  than~$\alpha_2$, it  ``epistemically precedes''
$\alpha_2$ in this run.

Theorem~\ref{thm:nodeNKgain}  shows 
that indeed the uneven centipede is necessary
for nb-nested knowledge gain. Interestingly, the proof
argument is identical to that which was used to prove the original Theorem~\ref{thm:NKgain}.
These proofs, originating in~\cite{BzMDISC2010}, can be found in the Appendix. 
The only thing changed is the formal language, and the proposed
communication patterns. The same goes for Theorem~\ref{thm:nodeCKgain} below, which utilizes 
the same proof as Theorem~\ref{thm:CKgain}, with an update from $\Langold$
to $\Langnew$.
It is a case where our understanding of what is going on has been 
limited purely by the expressivity of the formal apparatus of which we made use.

\begin{theorem}
\label{thm:nodeNKgain}
    Let~$r\in\Rrep$.
    Assume that $\trigg$ is an external event occurring at $\alpha_{0}=\node{i_0,t_0}$ in $r$.\\
    If $(\Rrep,r)\sat K_{\alpha_k}K_{\alpha_{k-1}}\!\!\cdots
    K_{\alpha_{0}}\tocc{t_0}{\trigg}$, 
    then there is an uneven centipede for $\langle \alpha_0,\ldots,\alpha_k\rangle$ in $r$.
\end{theorem}

We now turn to explore the implications of nb-nested knowledge 
for the coordinated Ordered Response problem. 
In line with our above discussion concerning nb-semntics and the decoupling
of epistemic necessity and past tense, we expect that nb-nested knowledge will allow
for greater flexibility in the timing of performed responses, solving tasks such as the following:

\emph{Once Charlie deposits the money, you will sign the 
contract, and I will deliver the merchandise no more than 
$5$ days after you sign.}

This suggests the   \ORi variant 
$\angles{deposit,\rsp{\node{You,sign}},\rsp{\node{I,deliver}}}$,
where the required timing is such that $t_{deliver}\leq t_{sign}+5$. Assume that
\emph{Charlie, You} and \emph{I} act at nodes $\alpha_{Ch},\alpha_{Y}$ and $\alpha_{I}$
respectively. By modifying Theorem~\ref{thm:ORnestedK}, we should be able to show
that \mbox{$K_{\alpha_{I}}K_{\alpha_{Y}}\tocc{t_{Y}}{deposit}$} holds. Once we have shown this,
Theorem~\ref{thm:nodeNKgain} will show that the required communication
is as shown in Figure~\ref{fig:wtr1}.

\begin{figure}[!h]
	\centering
	\fbox{
	\subfloat[]{\label{fig:wtr1}
		\newcommand{\tikzfname}{tikz_wtr1}
		\tikzsetnextfilename{Figures/\tikzfname}	
		\resizebox{0.45\linewidth}{!}{ \begin{tikzpicture}[font=\huge]
        \node[agent_time=\nodeC, label=left:{$\alpha_{0}$}]  (i0) at (0,0) {};
        	\coordinate (ct) at (4,1) {};
        	\coordinate (c1) at (6,2.4) {};
        	\coordinate (c2) at (6,1.7) {};
        	\coordinate (c3) at (6,1) {};
        	\coordinate (ck) at (6,-0.5) {};
        
        \path (i0) -- (ct) node[pos=0.9,agent_time=\nodeC,  label=above:{$\theta$}] (t)  {};   
        \path (t) -- (c1) node[pos=1.2,agent_time=\nodeC, label=right:{$\alpha_{1}$}] (i1)  {};   
        \path (t) -- (c2) node[pos=1,agent_time=\nodeC, label=right:{$\alpha_{2}$}] (i2)  {};   
        \path (t) -- (c3) node[pos=1.6,agent_time=\nodeC, label=right:{$\alpha_{3}$}] (i3)  {};   
        \path (t) -- (ck) node[pos=1.3,agent_time=\nodeC, label=right:{$\alpha_{k}$}] (ik)  {};

        \draw[loosely dotted, ultra thick, shorten <=6pt, shorten >=6pt] (c3) to (ck);

          \draw[syncausality] (i0) to (t);

           \draw[guarantee] (t) to (i1);
           \draw[guarantee] (t) to (i2);
           \draw[guarantee] (t) to (i3);
           \draw[guarantee] (t) to (ik);
 \end{tikzpicture}}
	}
	\subfloat[]{\label{fig:wtr2}
		\newcommand{\tikzfname}{tikz_wtr2}
		\tikzsetnextfilename{Figures/\tikzfname}
		\resizebox{0.45\linewidth}{!}{ \begin{tikzpicture}[font=\huge]
        \node[agent_time=\nodeC, label=left:{$\alpha_{0}$}]  (i0) at (0,0) {};
        	\coordinate (ct) at (4,1) {};
        	\coordinate (c1) at (6,2.4) {};
        	\coordinate (c2) at (6,1.7) {};
        	\coordinate (c3) at (6,1) {};
        	\coordinate (ck) at (6,-0.5) {};
        
        \path (i0) -- (ct) node[pos=0.9,agent_time=\nodeC,  label=above:{$\theta$}] (t)  {};   
        \path (t) -- (c1) node[pos=1.2,agent_time=\nodeC, label=right:{$\alpha_{1}$}] (i1)  {};   
        \path (t) -- (c2) node[pos=1,agent_time=\nodeC, label=right:{$\alpha_{2}$}] (i2)  {};   
        \path (t) -- (c3) node[pos=1.6,agent_time=\nodeC, label=right:{$\alpha_{3}$}] (i3)  {};   
        \path (t) -- (ck) node[pos=1.3,agent_time=\nodeC, label=right:{$\alpha_{k}$}] (ik)  {};

        \draw[loosely dotted, ultra thick, shorten <=6pt, shorten >=6pt] (c3) to (ck);

          \draw[syncausality] (i0) to (t);

           \draw[guarantee] (t) to (i1);
           \draw[guarantee] (t) to (i2);
           \draw[guarantee] (t) to (i3);
           \draw[guarantee] (t) to (ik);
 \end{tikzpicture}}
	}
	}
	\caption{} 
\end{figure}

Perhaps surprisingly, it turns out that the required epistemic state for ensuring the
correct execution of the problem as defined, is actually\\
\mbox{$K_{\alpha_{Y}}K_{\alpha_{I}}\tocc{t_{I}}{deposit}$!} Which in turn requires the 
inverted centipede shown in Figure~\ref{fig:wtr2}. To see why,  consider that the requirement
$t_{deliver}\leq t_{sign}+5$ again. Seemingly an upper bound constraint upon \emph{me} to deliver
the merchandise no more than $5$ days after contract sign, 
it can also be read as a lower bound constraint upon \emph{You}
to sign the contract no sooner than $5$ days before delivery takes place...
The point is, however, that \emph{I} cannot promise to deliver the merchandise within
the set time, since word of the signing may arrive at my site a week after
it occurs. Whereas \emph{You}, no matter how late (or how soon) you hear of my having delivered
the merchandise,  can always wait a few days to ensure that the $5$ days have passed.

The discussion concerning this timed coordination task points out 
a valuable insight. The knowledge that each  agent must possess \emph{as it responds}, 
concerns those responses whose time of occurrence is already bounded from above. 
These imply a lower bound for the agent's own action. In the 
first proposed (informal) definition, it is actually the case that \emph{You} have knowledge of an upper
limit upon \emph{my} response. This brings about the reversal of the required nested knowledge,
in comparison to what we expected it to be like. A task definition that exposes the 
implied \emph{upper bounds} is in place.

We now provide a formal definition for the kind of coordination task
that is characterized by nb-nested knowledge. The definition is phrased in 
accordance with the above discussion, and as Theorem~\ref{thm:WTRnestedK} below shows, 
solutions to the problem indeed require that the intuitively proper nested
knowledge holds. The required coordination is considered \emph{weakly timed},
reflecting the added existence of timing constraints, which only bound from below.
As we will see in the next section, the \emph{tightly timed} task is similar, but contains
stronger bounds on relative timing of responses.

\begin{definition}[Weakly Timed Response] \label{def:WTR}
	Let $\trigg$ be an external input nondeterministic event.
	A protocol~$P$ solves the instance\\
	$\WTRi=\angles{\trigg,\rsp{\alpha_1}\!:\delta_{1},\rsp{\alpha_{2}}\!:\delta_{2},\ldots,
	\rsp{\alpha_{k-1}}\!:\delta_{k-1},\rsp{\alpha_k}}$
	of the {\em Weakly Timed Response}
	problem if it guarantees that
	\begin{enumerate}
		\item every response $\rsp{\alpha_h}$, for $h=1,\ldots,k$, occurs in a run
		iff the trigger event $\rsp{\trigg}$ occurs in that run.
		\item in a run where response $\rsp{\alpha_{h}}$ occurs 
		at $\alpha_{h}=\node{i_{h},t_{h}}$ for all $h\leq k$,
		for every such $h$ we have that $t_{h+1} \geq t_{h}+\delta_{h}$. 
	\end{enumerate}
\end{definition}

Before stating the theorem relating the \WTR~problem to nb-nested knowledge,
another nuance should be observed. The problem definition specifies that even though
 agent $i_{k}$ may not know the exact time at which responses are performed by other agents,
it can work out an upper bound on the time of responses carried out by agents $i_{1}$ to $i_{k-1}$.
For example, response $\rsp{\alpha_{k-1}}$ gets carried out at $t_{k-1}$ which no later than 
$t_{k}-\delta_{k-1}$. Response $\rsp{\alpha_{k-2}}$ is then bounded with respect to $\alpha_{k}$ by 
$t_{k-2}\leq t_{k-1}-\delta_{k-2}\leq t_{k}-\delta_{k-1}-\delta_{k-2}$, etc. 
We will use 
$$\beta^{k}_{h}=\node{i_{h},t^{k}_{h}}\qquad \textrm{where }
t^{k}_{h} = t_{k}-\sum_{j=h}^{k-1}\delta_{j}$$
to denote this upper limit: the latest possible node at which response $\rsp{\alpha_{h}}$ gets
carried out, given that response $\rsp{\alpha_{k}}$ is performed at time $t_{k}$. Note that 
$t_{h}\leq t^{k}_{h}$ since by definition $t^{k}_{h}$ is an upper bound on $t_{h}$. For the same
reason we also have that $t^{k}_{h} \leq t^{k+1}_{h}$, since
$$t^{k+1}_{h} = t_{k+1}-\delta_{k}-\sum_{j=h}^{k-1}\delta_{j} = 
t^{k+1}_{k}-\sum_{j=h}^{k-1}\delta_{j}\geq t_{k}-\sum_{j=h}^{k-1}\delta_{j} = t^{k}_{h}.$$

\begin{theorem}
\label{thm:WTRnestedK}
	Let 
	$\WTRi=\angles{\trigg,\rsp{\alpha_1}\!:\delta_{1},\rsp{\alpha_{2}}\!:\delta_{2},\ldots,
	\rsp{\alpha_{k-1}}\!:\delta_{k-1},\rsp{\alpha_k}}$
	be an instance of \WTR,
	and assume that \WTRi~is solved in the system $\Rrep$.
	Let $r\in\Rrep$ be a run in which~$\trigg$ occurs.
	For each $h\leq k$, let $\alpha_h=\node{i_{h},t_{h}}$ be the
	node at which  response~$\rsp{\alpha_h}$ gets performed,
	and let 
	$$\beta^{k}_{h}=\node{i_{h},t_{k}-\sum_{j=h}^{k-1}\delta_{j}}.$$
	Then
	\[(\Rrep,r)\sat~K_{\alpha_{k}}K_{\beta^{k}_{k-1}}\cdots K_{\beta^{k}_{1}}\tocc{t_{1}}{\trigg}.\]
\end{theorem}

\section{Common Knowledge and Tight Bounds}\label{sec:TTR}
\begin{figure}[!h]
\centering
\fbox{
\newcommand{\tikzfname}{tikz_nbbroom}
\tikzsetnextfilename{Figures/\tikzfname}
\resizebox{0.6\linewidth}{!}{ \begin{tikzpicture}[font=\huge]
        \node[agent_time=\nodeC, label=left:{$\alpha_{0}$}]  (i0) at (0,0) {};
        	\coordinate (ct) at (4,1) {};
        	\coordinate (c1) at (6,2.4) {};
        	\coordinate (c2) at (6,1.7) {};
        	\coordinate (c3) at (6,1) {};
        	\coordinate (ck) at (6,-0.5) {};
        
        \path (i0) -- (ct) node[pos=0.9,agent_time=\nodeC,  label=above:{$\theta$}] (t)  {};   
        \path (t) -- (c1) node[pos=1.2,agent_time=\nodeC, label=right:{$\alpha_{1}$}] (i1)  {};   
        \path (t) -- (c2) node[pos=1,agent_time=\nodeC, label=right:{$\alpha_{2}$}] (i2)  {};   
        \path (t) -- (c3) node[pos=1.6,agent_time=\nodeC, label=right:{$\alpha_{3}$}] (i3)  {};   
        \path (t) -- (ck) node[pos=1.3,agent_time=\nodeC, label=right:{$\alpha_{k}$}] (ik)  {};

        \draw[loosely dotted, ultra thick, shorten <=6pt, shorten >=6pt] (c3) to (ck);

          \draw[syncausality] (i0) to (t);

           \draw[guarantee] (t) to (i1);
           \draw[guarantee] (t) to (i2);
           \draw[guarantee] (t) to (i3);
           \draw[guarantee] (t) to (ik);
 \end{tikzpicture}}
}
\caption{A uneven broom}
\label{fig:nb-broom}
\end{figure}
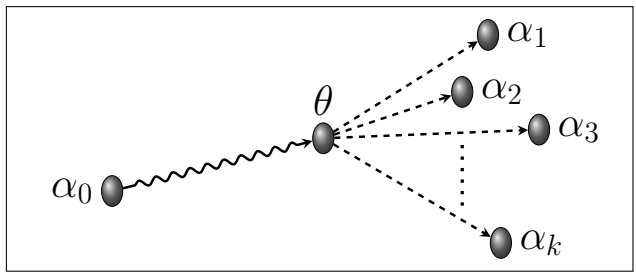

In this section we examine nb-common knowledge and 
its relation to communication and coordination. 
Just as nb-nested knowledge requires an extension to the centipede structure,
nb-common knowledge can only arise if the \emph{uneven broom} communication 
structure, seen in Figure~\ref{fig:nb-broom} and defined below, takes place in the run.
(``Uneven'' again comes from the broom's uneven legs.)

\begin{definition}[Uneven broom]
	\label{def:broom}
	Let $A=\{\alpha_{1,\ldots,\alpha_{k}}\}$ be a sequence of nodes 
	and let $\alpha_{0}$ be a node.
	Node $\theta$ is an {\em (uneven) broom} for  $\angles{\alpha_{0},A}$ in $r$ if
	~$\alpha_{0}\spc\theta$ ~and ~$\theta\tspc \alpha_{h}$ holds for
        all ~$h=1,..,k$. 
\end{definition}

Theorem~\ref{thm:nodeCKgain} proves that the uneven broom characterizes 
the necessary communication for nb-common knowledge to arise.

\begin{theorem}
	\label{thm:nodeCKgain}
	Let~ $A\subseteq\CV$ with $\node{i_{k},t_{k}}$ being the latest node 
	($t_{k}\geq t_{h}$ for all $\node{i_{h},t_{h}}\in A$) , and let $r\in\Rrep$. Assume
	that   $\trigg$ is an external input event at $\alpha_{0}$ in~$r$. 
	If $(\Rrep,r)\sat C_A(\tocc{t_{k}}{\trigg})$,
	then there is a  uneven broom $\hat \theta$ for $\angles{ \alpha_0,A}$ in
	$r$.
\end{theorem}

Recall
the Simultaneous Response problem, that shows standard common knowledge 
as the necessary requirement when  a group of agents need to act simultaneously.
Yet, as discussed in the introduction, node based common knowledge does  away
with simultaneity while retaining other properties of common knowledge. 
In accordance, we expect it to serve as the epistemic requirement for some weakened generalization of the
simultaneity requirement. 
The following coordination task seems like a promising candidate.

\begin{definition}[Tightly Timed Response]
	Let $\trigg$ be an external input nondeterministic event.
	A protocol~$P$ solves the instance
	$$\TTRi=\angles{\trigg,\rsp{\alpha_1}\!:\delta_{1},\rsp{\alpha_2}\!:\delta_{2},
	\ldots,\rsp{\alpha_k}\!:\delta_{k}}$$ 
	of the {\em Tightly Timed Response} 
	problem if it guarantees that
	\begin{enumerate}
		\item every response $\rsp{\alpha_h}$, for $h=1,\ldots,k$, occurs in a run
		iff the trigger event $\trigg$ occurs in that run.
		\item For every $h,g\leq k$
		the relative timing of the responses is exactly the difference 
		in the associated delta values: $t_{h}-t_{g} = \delta_{h}-\delta_{g}$
	\end{enumerate}
\end{definition}

Note how the new  $\TTR$ problem definition generalizes the previous $\SR$ definition:
A simultaneous response problem  $\SRi=\angles{\trigg,\rsp{\alpha_1},\ldots,\rsp{\alpha_k}}$ is simply 
an extremely tight $\TTR$ problem, where $\delta_{1},..,\delta_{k}$ are all set to $0$.

Intuitively, we expect that an agent $i$ participating in a solution to a $\TTR$ task by
performing response $\alpha_{i}$ at time $t_{i}$ will know that agent $j$ will perform
its own response $\alpha_{j}$ at precisely $t_{j} = \delta_{j}-\delta_{i}+t_{i}$. We
also expect agent $j$ at $t_{j}$ to know that $i$ is carrying out $\alpha_{i}$ at $t_{i}$.
Theorem~\ref{thm:TTRck} below shows that these individually specified knowledge states,
and others that are derived from the $\TTR$ definition, add up to a node based
common knowledge gain requirement. 

\begin{theorem}\label{thm:TTRck}
	Let $\TTRi=\TTRi=\angles{\trigg,\rsp{\alpha_1}\!:\delta_{1},\rsp{\alpha_2}\!:\delta_{2},
	\ldots,\rsp{\alpha_k}\!:\delta_{k}}$,
	and assume that $\TTRi$ is solved in~$\Rrep$.
	Let $r\in\Rrep$ be a run in which~$\trigg$ occurs, and let 
	$A=\{\alpha_1,\ldots,\alpha_k\}$ be the set of nodes 
	at which the responses are carried out in the run $r$ ($\rsp{\alpha_{1}}$ occurs at node 
	$\alpha_{1}$, etc.). WLOG, let $\alpha_{h}=\node{i_{h},t_{h}}$ be the earliest node in $A$.
	Then ~~$(\Rrep,r)\sat~C_A \tocc{t_{h}}{\trigg}.$
\end{theorem}

Summing up, the $\TTR$ problem requires agents to be fully informed with
respect to each other's response time - without the extra requirement for simultaneity.
Since each of the agents, as it responds, knows the response times of all other agents - we end
up with node based common knowledge, which in turn still requires  that a broom node exist
that can guarantee that all of its messages to the responding agents are delivered by the time these
agents are set to respond. 

\section{Conclusions}\label{sec:conc}
This paper explores the implications of a new formalism for epistemic statements
upon the notions of nested and common knowledge. It checks how these altered concepts
interact with communication on the one hand, and coordination on the other - along the lines
of~\cite{BzMDISC2010,BenZvi2011}. The new formalism is seen to allow 
for a decoupling of epistemic necessity and the past tense,
in the sense that knowing that an event must occur or that an agent gains knowledge of a fact
no longer entails that the occurrence, or the knowledge gain, have happened in the past.
This, in turn, allows for a wider range of coordination tasks to be characterized in terms of the epistemic
states that they necessitate. We define two such coordination tasks. The Weakly Timed and the
Tightly Timed Response problems, extending our previous definitions of Ordered and Simultaneous 
Response~\cite{BzMDISC2010,BenZvi2011}. 

Our analysis of the relations between knowledge and coordination yields  two valuable
insights. First, when agents must respond within interrelated time bounds, as seen in the
 \WTR~problem, the crucial knowledge that they must gain before applying their assigned
 actions concerns those responses whose response times are \emph{bounded from above}
 with respect to their own response. This allows them, if necessary, to delay their own action
in order to conform with the problem requirements. 

The enquiry into the node-based extension to common knowledge is even more rewarding.
Much has been written about the relation between common knowledge and simultaneity.
Formal analysis of this relation is given in~\cite{HM1,FHMV}. Other analyses implicitly
rely upon simultaneity in accounting for common knowledge gain, using such concepts as
\emph{public announcements}~\cite{Baltag99} or the \emph{copresence heuristics}~\cite{CM81}. 
The analysis presented here sheds light on this issue by generalizing common knowledge
and showing that the more general form corresponds to a temporally tight form of coordination. 
The previously established connection between common knowledge and simultaneity is, 
in fact, a particular instance of this more general connection. 
Recall by Lemma~\ref{lem:timestamping} that there is an embedding of the standard common knowledge 
$C_G\occurred{\trigg}$ at time~$t$ into the formula $C_A\tocc{t}{\trigg}\in\Langnew$
where all nodes in $A$ are timed to $t$. 
The particular form of coordination corresponding to the embedded formula 
is tight coordination with all deltas set to $0$---namely, simultaneity at time~$t$.

We consider this paper as a point of departure for several lines of potential further research.
First, we have only touched the surface of the required logical analysis for node
based semantics. Completeness and tractability issues remain unknown, as well as
possible variations on the node based theme that will chime in with 
ongoing research in the temporal logic and model checking communities.
Second, although we have mainly been concerned with temporally oriented
coordination, the analysis may also be instructive where 
other aspects of coordination are concerned. Chwe's~\cite{Chwe2000} 
analysis of the communication
network required in order to bring about a revolt in a social setting 
is a case in point for further research.
Finally, we have yet to study the implications of the \WTR~and \TTR~problems defined 
here in the context of distributed computing tasks, where timed coordination is 
often essential.

\bibliography{z}
\newpage

\appendix
\section{Proofs}\label{app}
This appendix contains all lemmas and major proofs necessary to support
the claims of the paper. Note that the formal work leading up to Theorems~\ref{thm:nodeNKgain}
and~\ref{thm:nodeCKgain} is non-trivial, and the interested reader would be better off reading
the more detailed account presented in~\cite{BzMDISC2010,BzMLOFT2010,BenZvi2011}.

\renewcommand{\toto}{lem:timestamping}
\begin{rlemma}
	There exists a function $\ts: \Langold \times Time \to \Langnew$
	such that for every $\vphi\in\Langold$ , time $t$, and $\vphi^{t}=\ts(\vphi,t)$:
	$$(\Rrep,r,t)\sat \vphi \qquad \textrm{iff} \qquad (\Rrep,r)\sat \vphi^{t}.$$
\end{rlemma}

\begin{proofL} 
	Proof is by structural induction on $\vphi$. We go over the clauses. Each clause 
	is used to define $\ts$, but it is also easy to see that based on the
	clause definition $(\Rrep,r,t)\sat \vphi$ iff $(\Rrep,r)\sat \vphi^{t}$.
	\begin{itemize}
		\item $\ts( \occurred{e},t) = \tocc{t}{e}$.
		\item $\ts( K_{i}\psi ,t) = K_{\node{i,t}}\psi$.
		\item $\ts( E_{G}\psi, t) =  E_{A}\psi$ where 
		$A = \{\node{j,t} : j\in G\}$.
		\item $\ts( C_{G}\psi,t) =  C_{A}\psi$  where 
		$A = \{\node{j,t} : j\in G\}$.
	\end{itemize}
\end{proofL}

\renewcommand{\toto}{lem:prop-nodeCK}
\begin{rlemma}\mbox{}
	\begin{itemize}
		\item Both $K_{\node{i,t}}$ and $C_A$ are $S5$ modalities.
		\item $C_{A}\vphi \imp E_{A}(\vphi \wedge C_{A}\vphi)$ is valid.
		\item If ~$\Rrep\sat \vphi \imp E_{A}(\vphi \wedge \psi)$ ~then ~$\Rrep\sat \vphi \imp C_{A}\psi$
	\end{itemize}
\end{rlemma}

\begin{proofL}
	\begin{description}
		\item For $S5$ modality the $K_{\node{i,t}}$ case is immediate.
			We focus on showing for $C_{A}$:
			\begin{description}
			\item[\textbf{K \; ($C_{A}\vphi \wedge C_{A}(\vphi \imp \psi) \imp C_{A}\psi$):}] 
			Suppose that $(\Rrep,r)\sat C_{A}\vphi \wedge C_{A}(\vphi \imp \psi) \wedge \neg C_{A}\psi$ for some
			$r,\vphi,\psi$ and $A$. Then there is a sequence $\alpha_{1},\alpha_{2},..,\alpha_{n}\in A$
			such that $(\Rrep,r)\nsat K_{\alpha_{n}}K_{\alpha_{n-1}}\cdots K_{\alpha_{1}}\psi$.
			From $(\Rrep,r)\sat C_{A}\vphi$ and $(\Rrep,r)\sat C_{A}(\vphi \imp \psi)$ we obtain
			$(\Rrep,r)\sat K_{\alpha_{n}}K_{\alpha_{n-1}}\cdots K_{\alpha_{1}}\vphi$ and
			$(\Rrep,r)\sat K_{\alpha_{n}}K_{\alpha_{n-1}}\cdots K_{\alpha_{1}}(\vphi \imp \psi)$
			respectively,  and using the \textbf{K} axiom for the knowledge operator and trivial
			induction we get that  
			$(\Rrep,r)\sat K_{\alpha_{h}}K_{\alpha_{h-1}}\cdots K_{\alpha_{1}}\psi$ for all $h\leq n$, 
			and in particular
			for $h=n$. This contradicts the assumption that 
			$(\Rrep,r)\sat C_{A}\vphi \wedge C_{A}(\vphi \imp \psi) \wedge \neg C_{A}\psi$.
			
			\item[\textbf{T \;($C_{A}\vphi \imp \vphi)$:}] 
			Suppose that $(\Rrep,r)\sat C_{A}\vphi$ for some
			$r,\vphi$ and $A$. Fix some $\node{i,t}\in A$. From  $(\Rrep,r)\sat C_{A}\vphi$ we get
			$(\Rrep,r)\sat K_{\node{i,t}}\vphi$, and by definition of $\sat$ we conclude that 
			$(\Rrep,r')\sat \vphi$ for all $r'$ such that $r'_{i}(t)=r_{i}(t)$, and in particular for $r'=r$.
			Thus  $(\Rrep,r)\sat\vphi$.
			
			\item[\textbf{4 \;($C_{A}\vphi \imp C_{A}C_{A}\vphi)$:}] Suppose not. Then there exist
			$r,A,\vphi$ such that  $(\Rrep,r)\sat C_{A}\vphi$ and a sequence $\alpha_{1},\alpha_{2},..,\alpha_{n}\in A$
			such that  $(\Rrep,r)\nsat K_{\alpha_{n}}K_{\alpha_{n-1}}\cdots K_{\alpha_{1}}C_{A}\vphi$.
			Once again, there must be a sequence $\alpha'_{1},\alpha'_{2},..,\alpha'_{m}\in A$ such that
			$$(\Rrep,r)\nsat K_{\alpha_{n}}\cdots K_{\alpha_{1}}K_{\alpha'_{m}}\cdots K_{\alpha'_{1}}\vphi.$$
			Yet this contradicts $(\Rrep,r)\sat C_{A}\vphi$.
			
			\item[\textbf{5 \;($\neg C_{A}\vphi \imp C_{A}\neg C_{A}\vphi)$:}] Suppose not.	Then there exist
			$r,A,\vphi$ such that  $(\Rrep,r)\sat \neg C_{A}\vphi$ and a sequence 
			$\alpha_{1},\alpha_{2},..,\alpha_{n}\in A$
			such that  $(\Rrep,r)\nsat K_{\alpha_{n}}K_{\alpha_{n-1}}\cdots K_{\alpha_{1}}\neg C_{A}\vphi$.
			By definition of $\sat$ there exists $r'$ s.t. $r^{n}_{i_{n}}(t_{n})=r_{i_{n}}(t_{n})$ and where
			\[(\Rrep,r^{n})\nsat K_{\alpha_{n-1}}K_{\alpha_{n-1}}\cdots K_{\alpha_{1}}\neg C_{A}\vphi.\]
			Similar argumentation provides us with runs $r^{h}$ for all $1\leq h<n$ such that
			$(\Rrep,r^{h})\nsat K_{\alpha_{h-1}}K_{\alpha_{h-2}}\cdots K_{\alpha_{1}}\neg C_{A}\vphi$.
			In particular for $h=1$ we get that
			$(\Rrep,r^{1})\nsat \neg C_{A}\vphi$, and hence $(\Rrep,r^{1})\sat C_{A}\vphi$, and finally that 	
			$(\Rrep,r^{1})\sat K_{\alpha_{n}}K_{\alpha_{n-1}}\cdots K_{\alpha_{1}}C_{A}\vphi$. Yet by build we have 
			(i) $r^{1}_{i_{1}}(t_{1})=r^{2}_{i_{1}}(t_{1})$, $r^{2}_{i_{2}}(t_{2})=r^{3}_{i_{2}}(t_{2})$, etc., giving
			us  $(\Rrep,r)\sat C_{A}\vphi$, contradicting our assumption.
			
			\item[\textbf{N \;(If $\sat \vphi$ then $\sat C_{A}\vphi)$:}] We show by induction on $h$
			that for any $h$ length sequence $\alpha_{1},\alpha_{2},..,\alpha_{h}\in A$ we have
			$\sat K_{\alpha_{h}}K_{\alpha_{h-1}}\cdots K_{\alpha_{1}}\vphi$. For $h=0$ this is immediate
			from the assumption. Assume for $h$. Fix run $r$. Note that for every run $r'$ s.t. 
			$r_{i_{h}}(t_{h})=r'_{i_{h}}(t_{h})$ we have 
			$(\Rrep,r')\sat K_{\alpha_{h-1}}\cdots K_{\alpha_{1}}\vphi$ by the inductive hypothesis. Hence
			we get that $(\Rrep,r)\sat K_{\alpha_{h}}K_{\alpha_{h-1}}\cdots K_{\alpha_{1}}\vphi$. As this is true
			for any $r$ we conclude that $\sat K_{\alpha_{h}}K_{\alpha_{h-1}}\cdots K_{\alpha_{1}}\vphi$
			as required.
			By definition of $C_{A}\vphi$ we get that  $\sat C_{A}\vphi$.
		\end{description}•
		 
		 \item[$C_{A}\vphi \imp E_{A}(\vphi \wedge C_{A}\vphi)$:]   Fix $r,\vphi,A$. 
		 From $(\Rrep,r)\sat C_{A}\vphi$ we get by definition of $C_{A}$ 
		 that (i)  $(\Rrep,r)\sat K_{\alpha}\vphi$ and (ii) $(\Rrep,r)\sat K_{\alpha}C_{A}\vphi$ 
		 for all $\alpha\in A$. Hence we also get 
		 (iii) $(\Rrep,r)\sat E_{A}\vphi$ and (iv) $(\Rrep,r)\sat E_{A}C_{A}\vphi$. Putting (iii) and (iv)
		 together we get $(\Rrep,r)\sat E_{A}(\vphi \wedge C_{A}\vphi)$.
		 
		 \item[If $\Rrep\sat \vphi \imp E_{A}(\vphi \wedge \psi)$ then $\Rrep\sat \vphi \imp C_{A}\psi$:] 
		 Suppose \mbox{$\Rrep\sat \vphi \imp E_{A}(\vphi \wedge \psi)$.} Fix a sequence 
		 $\alpha_{1},\alpha_{2},..,\alpha_{k}\in A$. We will now show by induction
		 on $h\leq k$that 
		 \mbox{$\Rrep\sat \vphi \imp K_{\alpha_{k}}K_{\alpha_{k-1}}\cdots K_{\alpha_{1}}\psi$}. 
		 For $h=1$ we have by assumption  that $\Rrep\sat \vphi \imp E_{A}(\vphi \wedge \psi)$. 
		 We replace $E_{A}$ by $K_{\alpha_{1}}$, recalling that $\alpha_{1}\in A$.
		 We get  $\Rrep\sat \vphi \imp K_{\alpha_{1}}\psi$. Assume for $h$ and look at the case of $h+1$.
		 By assumption $\Rrep\sat \vphi \imp E_{A}(\vphi \wedge \psi)$  and hence 
		 $\Rrep\sat \vphi \imp K_{\alpha_{h+1}}(\vphi \wedge \psi)$. Fix run $r$. If $(\Rrep,r)\nsat \vphi$
		 then $(\Rrep,r)\sat \vphi \imp K_{\alpha_{h+1}}K_{\alpha_{h}}\cdots K_{\alpha_{1}}\psi$ and we are done.
		 Else, we have that $(\Rrep,r)\sat E_{A}(\vphi \wedge \psi)$, and so 
		 $(\Rrep,r)\sat K_{\alpha_{h+1}}(\vphi \wedge \psi)$. Thus, we get that 
		 $(\Rrep,r')\sat \vphi$ for every $r'$ such that $r_{i_{h+1}}(t_{h+1})=r'_{i_{h+1}}(t_{h+1})$.
		 By the inductive hypothesis we get that $(\Rrep,r')\sat K_{\alpha_{h}}\cdots K_{\alpha_{1}}\psi$.
		 By definition of $\sat$ we get that $(\Rrep,r)\sat K_{\alpha_{h+1}}K_{\alpha_{h}}\cdots K_{\alpha_{1}}\psi$.
		 We conclude that $\Rrep\sat \vphi \imp K_{\alpha_{k}}K_{\alpha_{k-1}}\cdots K_{\alpha_{1}}\psi$.
		 As this holds for any sequence of nodes in $A$ we get that 
		 $\Rrep\sat \vphi \imp C_{A}\psi$ as required.
	\end{description}
\end{proofL}

\begin{definition}[Past cone]
	Fix $r\in\Rrep$ and $\theta\in\CV$. The past cone of $\theta$ in $r$, $\pas(r,\theta)$,
	is the set $\{\psi : \psi \spc \theta\}$.
\end{definition}

\begin{definition}[Agree upon]
We say that two runs $r$ and~$r'$ {\em agree on the node~$(i,t)\in\CV$} if 
\begin{enumerate}
	\item $r_i(t)=r'_i(t)$, 
	\item the same external inputs and messages arrive at $(i,t)$ in both runs, and 
	\item the same actions are performed by~$i$ at time~$t$. 
\end{enumerate}
\end{definition}

\begin{lemma}\label{lem:past-spontaneous}
Let $r\in\Rrep$ and let $\theta\in\CV$. 
Then there is a run~$r'\in\Rrep$ such that 
\begin{enumerate}
	\item $\pas(r',\theta)=\pas(r,\theta)$, 
	\item $r'$ and~$r$ agree on all the nodes of $\pas(r',\theta)$. 
	\item the only nondeterministic events in~$r'\!$ occur at nodes 
	of $\pas(r',\theta)$. 
\end{enumerate}
\end{lemma}

\begin{lemma}\label{lem:step-synch}
If $\node{i,t}\spc \node{j,t'}$ then $t\le t'$, with $t=t'$ holding only if
$i=j$.
\end{lemma}

\begin{definition}[Early delivery]
	When a message sent at $\node{i,t}$ arrives at $\node{j,t'}$
	prior to the maximal allowed delay, i.e. when
	$t'<t+\bij$, then we say that an early delivery nondeterministic event has occurred.
\end{definition}

\begin{lemma}\label{lem:bridges-exist}
Fix a run~$r$, and let $\theta\spc\theta'$. If $\theta\not\tspc\theta'$
then there is a node~$\beta$  
such that $\theta\spc\beta\tspc\theta'$ and an  early delivery occurs
at~$\beta$ in~$r$. 
\end{lemma}

\begin{lemma}\label{thm:2proc}
Let $e$ be the delivery of an external input that occurs at the node
$\alpha_{0}$ in~$r\in\Rrep$.
If ~$(\Rrep,r)\sat K_{\alpha_{1}}\tocc{t_{1}}{e}$
then $\alpha_{0}\spc \alpha_{1}$.
\end{lemma}

\renewcommand{\toto}{thm:nodeNKgain}
\begin{rtheorem}[Knowledge Gain]
    Let~$r\in\Rrep$.
    Assume that $e$ is an external input event occurring at $\alpha_{0}$ in $r$.\\
    If $(\Rrep,r)\sat K_{\alpha_k}K_{\alpha_{k-1}}\!\!\cdots
    K_{\alpha_{1}}\tocc{t_{1}}{e}$, 
    then there is an uneven centipede for $\langle \alpha_0,\ldots,\alpha_k\rangle$ in $r$.
\end{rtheorem}

\begin{proofT}
First note that  $k\ge 1$ and $(\Rrep,r)\sat
K_{\alpha_{k}}K_{\alpha_{k-1}}\cdots K_{\alpha_1}\tocc{t_{1}}{e}$ imply by the
Knowledge Axiom that $(\Rrep,r)\sat K_{\alpha_k}\tocc{t_{1}}{e}$. It follows
that $\alpha_{0}\spc\alpha_{k}$ in~$r$ by Lemma~\ref{thm:2proc}. We prove the
claim by induction on $k\ge 1$:

\begin{description} \item[$k=1$]\quad As argued above,
$\alpha_{0}\spc\alpha_{k}$. Thus, $\alpha_{0}\spc\alpha_{1}$ since $k=1$, and so
$\langle\alpha_{0},\alpha_{k}\rangle$ is a (trivial) centipede for $\langle
\alpha_{0},\alpha_{k} \rangle$ in $r$.

\item[$k\ge 2$]\quad Assume inductively that the claim holds for $k-1$.
Moreover, assume that $(\Rrep,r)\sat K_{\alpha_k}K_{\alpha_{k-1}}\cdots
K_{\alpha_1}\tocc{t_{1}}{e}$. Let~$r'$ be the run guaranteed by
Lemma~\ref{lem:past-spontaneous} to exist with respect to $r$, $i_k$
and~$t_{k}$. Recall from Lemma~\ref{lem:past-spontaneous}(1) that
$\pas(r',\alpha_{k})=\pas(r,\alpha_{k})$. Thus,  $\alpha_{0}\spc \alpha_{k}$ in
$r$ implies that $\alpha_{0}\spc \alpha_{k}$ in $r'$ too. Moreover, by
Lemma~\ref{lem:past-spontaneous}(2) we have that $r$ and $r'$ agree on the nodes
of $\pas(r,\alpha_{k})$, so in particular  $r'_{i_{k}}(t_{k})=r_{i_k}(t_{k})$.
Since $(\Rrep,r)\sat K_{\alpha_{k}}K_{\alpha_{k-1}}\cdots
K_{\alpha_1}\tocc{t_{1}}{e}$ and $r'_{i_{k}}(t_{k})=r_{i_k}(t_{k})$, we have
that \[(\Rrep,r',t')\sat K_{\alpha_{k-1}}\cdots K_{\alpha_1}\tocc{t_{1}}{e}.\]
By the inductive hypothesis there exists a centipede
$\langle\alpha_{0},\theta_1,\ldots,\theta_{k-1}\rangle$ for $\langle
\alpha_0,\ldots,\alpha_{k-1}\rangle$ in $r'$. Let $c\ge 0$ be the minimal index
for which $\theta_c\tspc\theta_h$ for all $h=c+1,\ldots,k-1$.  Clearly $c\le
k-1$, since ~$\theta_{k-1}\tspc\theta_{k-1}$.
\begin{itemize}
	\item If $c=0$ then 
	$\alpha_{0}\tspc \theta_h\tspc \alpha_{h}$, and thus also
	$\alpha_{0}\tspc \alpha_{h}$, 
	for $h=1,\ldots,k-1$. Since $\alpha_{0}\spc\alpha_{k}$ in $r$, it follows that 
	the tuple $\langle(\alpha_{0})^{d-1},\alpha_{k}\rangle$ (in which 
	$\alpha_{0}$ plays the role of the first $k-1$ nodes) 
	is a centipede for
	$\langle\alpha_{0},\ldots,\alpha_{k}\rangle$ in 
	    $r$.

	\item 
	Otherwise, $c>0$ and $\theta_{c-1}\spc\theta_c$ while 
	$\theta_{c-1}\not\tspc\theta_c$.  
	By Lemma~\ref{lem:bridges-exist} it follows that there exists a
	node~$\beta$ such that  $\theta_{c-1}\spc\beta\tspc\theta_c$, and 
	$\beta$ is the site of an early receive
	in the run~$r'$.  
	By construction of~$r'$, early receives can arrive only at nodes in
	$\pas(r',\alpha_{k})$.  
	It follows that $\beta\spc\alpha_{k}$ in $r'$, and since
	$\pas(r,\alpha_{k})=\pas(r',\alpha_{k})$, we have that  
	$\beta\spc\alpha_{k}$ in $r$ too and hence also that $\pas(r,\beta)=\pas(r',\beta)$. 
	It follows that in $r$
	\[\alpha_{0}\spc\theta_1\spc\cdots\spc\theta_{c-1}\spc\beta\spc\alpha_{k}.\]   
	Recall that $\tspc$ depends only on the weighted communication network, which
	is the same in both~$r$ and~$r'$.  
	Thus, $\theta_j\tspc\alpha_{j}$ for all $0<j\le c-1$. Moreover,
	$\beta\tspc\theta_h\tspc\alpha_{h}$ and so  
	$\beta\tspc\alpha_{h}$ 
	for all $h=c,c+1,\ldots,k-1$. 
	It follows that
	$\langle\alpha_{0},\theta_{1},\ldots,\theta_{c-1},(\beta)^{k-c},\alpha_{k}\rangle$ is a
	centipede  for $\langle\alpha_{0},\ldots,\alpha_{k}\rangle$ in $r$.
\end{itemize}

It follows that a centipede for $\langle \alpha_{0},\ldots,\alpha_{k}\rangle$ in 
$r$ is guaranteed to exist in all cases, 
as claimed.
\end{description}
\end{proofT}

\renewcommand{\toto}{thm:WTRnestedK}
\begin{rtheorem}
	Let \WTRi=$\angles{\trigg,\rsp{\alpha_1}\!:\delta_{1},\ldots,
	\rsp{\alpha_{k-1}}\!:\delta_{k-1},\rsp{\alpha_k}}$ be an instance of \WTR,
	and assume that \WTRi~is solved in the system $\Rrep$.
	Let $r\in\Rrep$ be a run in which~$\trigg$ occurs.
	For each $h\leq k$, let $\alpha_h=\node{i_{h},t_{h}}$ be the
	node at which  response~$\rsp{\alpha_h}$ gets performed,
	and let 
	$$\beta^{k}_{h}=\node{i_{h},t_{k}-\sum_{j=h}^{k-1}\delta_{j}}.$$
	Then
	\[(\Rrep,r)\sat~K_{\alpha_{k}}K_{\beta^{k}_{k-1}}\cdots K_{\beta^{k}_{1}}\tocc{t^{k}_{1}}{\trigg}.\]
\end{rtheorem}

\begin{proofT}
	We prove the theorem by induction on $h\leq k$.
	\begin{itemize}
		\item[] $\mathbf{h=1:}$ By definition of $r$, $\rsp{\alpha_1}$ gets performed at $\alpha_{1}$.
		Since performing a local action is written in the agent's local state (and hence known to the agent),  
		and since it is always performed no sooner than the triggering event $\trigg$,
		we have $(\Rrep,r)\sat K_{\alpha_{1}}\tocc{t_{1}}{\trigg}$. If $h=k=1$ then we are done.
		Else, as $t^{k}_{1}\geq t_{1}$ and as agents
		have perfect recall, we get	 $(\Rrep,r)\sat K_{\beta^{k}_{1}}\tocc{t^{k}_{1}}{\trigg}$.	

		\item[] $\mathbf{h>1:}$ Assume that 
		$(\Rrep,r)\sat~K_{\beta^{k}_{h-1}}\cdots K_{\beta^{k}_{1}}\tocc{t^{k}_{1}}{\trigg}$. 
		As response $\rsp{\alpha_{h}}$
		gets performed at $\alpha_{h}$ in $r$ we have $(\Rrep,r)\sat K_{\alpha_{h}}\tocc{t^{h}_{h}}{a_{h}}$.
		Since \WTRi~is solved in the system $\Rrep$ we have that in every run $r'$ such that 
		$r_{i_{h}}(t_{h})=r'_{i_{h}}(t_{h})$  response $\alpha_{h-1}$ gets performed at some $t_{h-1}$
		such that $t_{h-1}\leq t^{h}_{h-1}$. Note that every system that solves \WTRi~also solves the sub
		problem 
		 \WTRi'=$\angles{\trigg,\rsp{\alpha_1}\!:\delta_{1},\ldots,
		 \rsp{\alpha_{h-2}}\!:\delta_{h-2},\rsp{\alpha_{h-1}}}$. 
		 Based on the inductive hypothesis
		 we obtain that  
		 $$(\Rrep,r')\sat~K_{\alpha_{h-1}}K_{\beta^{k}_{h-2}}\cdots K_{\beta^{k}_{1}}\tocc{t^{k}_{1}}{\trigg}.$$
		 From $t^{k}_{h-1} \geq t_{h-1}$, and based on perfect recall, we derive that		  
		 $$(\Rrep,r')\sat~K_{\beta^{k}_{h-1}}K_{\beta^{k}_{h-2}}\cdots K_{\beta^{k}_{1}}\tocc{t^{k}_{1}}{\trigg}.$$
		 By our choice of runs $r'$ and the definition of $\sat$ this gives us
		 $$(\Rrep,r')\sat~K_{\alpha_{h}}K_{\beta^{k}_{h-1}}K_{\beta^{k}_{h-2}}
		 \cdots K_{\beta^{k}_{1}}\tocc{t^{k}_{1}}{\trigg}.$$
		 If $h=k$ then we are done. Else, once again deploying $t^{k}_{h} \geq t_{h}$ we conclude that
		 $$(\Rrep,r')\sat~K_{\beta^{k}_{h}}K_{\beta^{k}_{h-1}}K_{\beta^{k}_{h-2}}
		 \cdots K_{\beta^{k}_{1}}\tocc{t^{k}_{1}}{\trigg},$$
		 and we are done.
	\end{itemize}
\end{proofT}

\renewcommand{\toto}{thm:nodeCKgain}
\begin{rtheorem}
	Let~ $A\subseteq\CV$ with $\node{j_{k},t'_{k}}$ being the earliest node 
	($t'_{k}\leq t'_{h}$ for all $\node{j_{h},t'_{h}}\in A$) , and let $r\in\Rrep$. Assume
	that   $e$ is an external input event at $\alpha_{0}$ in~$r$. 
	If $(\Rrep,r)\sat C_A(\tocc{t'_{k}}{e})$,
	then there is a  uneven broom $\hat \theta$ for $\angles{ \alpha_0,A}$ in
	$r$.
\end{rtheorem}

\begin{proofT}
	Assume the notations and conditions of the theorem. Denote
	$A=\{\alpha_{1},\ldots,\alpha_{k}\}$ and $d=t'_{k}-t_{0}$, the time difference
	between the occurrence of $e$ and the latest node in $A$ . 
	Since $(\Rrep,r)\sat C_A (\tocc{t'_{k}}{e})$
	we have by definition of common knowledge that\\
	$(\Rrep,r)\sat E^{k(d+1)}_A\tocc{t'_{k}}{e}$. In particular, this implies that \[(\Rrep,r)\sat
	~(K_{\alpha_{k}}\cdots K_{\alpha_1})^{d+1}\tocc{t'_{k}}{e},\] where $(K_{\alpha_k}\cdots
	K_{\alpha_1})^{d+1}$ stands for~$d+1$ consecutive copies of $K_{\alpha_k}\cdots
	K_{\alpha_1}$. By the Knowledge Gain Theorem~\ref{thm:nodeNKgain}, there is a
	corresponding centipede
	$\sigma=\langle\theta_0,\theta_1,\ldots,\theta_{k(d+1)}\rangle$ in $r$.
	Denote $\theta_h=(i_h,t_h)$ for all $0\le h\le k\!\cdot\!(d+1)$. Recall
	that, by definition, $\theta_h\spc\theta_{h+1}$ holds for all
	$h<k\!\cdot\!(d+1)$. By Lemma~\ref{lem:step-synch} we obtain that if
	$\theta_h\ne\theta_{h+1}$ then $t_h<t_{h+1}$. It follows that there can be
	at most~$d+1$ distinct nodes $\theta'_1\spc\theta'_2\spc\cdots\spc\theta'\ell$
	in~$\sigma$. Every $\theta'_h$ represents a segment
	$\theta_x,\ldots,\theta_{x+s}$ of the nodes in~$\sigma$. By the pigeonhole
	principle, one of the $\theta'$ nodes must represent a segment consisting of at
	least~$k$ of the original $\theta$s in~$\sigma$. Denoting this node by~$\hat\theta$,
	we obtain that $\hat\theta\tspc \alpha_{h}$ for every $\alpha_{h}\in A$. Moreover, by
	definition of the centipede and transitivity of~$\spc$ we have that
	$\alpha_{0}\spc\hat\theta$. It follows that $\hat\theta$ is a centibroom 
	for $\langle \alpha_0,G\rangle$ in $r$.	
\end{proofT}

\renewcommand{\toto}{thm:TTRck}
\begin{rtheorem}
	Let $\TTRi=\angles{\trigg,\rsp{\alpha_1}\!:\delta_{1},\ldots,\rsp{\alpha_k}\!:\delta_{k}}$,
	and assume that $\TTRi$ is solved in~$\Rrep$.
	Let $r\in\Rrep$ be a run in which~$\trigg$ occurs, and let 
	$A=\{\alpha_1,\ldots,\alpha_k\}$ be the set of nodes 
	at which the responses are carried out in the run $r$ ($\rsp{\alpha_{1}}$ occurs at node 
	$\alpha_{1}=\node{i_{1},t_{1}}$, etc.). Let $\alpha'=\node{i',t'}$ be the earliest node in $A$.
	Then ~~$(\Rrep,r)\sat~C_A \tocc{t'}{\trigg}.$
\end{rtheorem}

\begin{proofT}
	Fix $h,g\in\{1..k\}$. We first show that 
	$$\Rrep\sat \tocc{t_{h}}{a_{h}}\imp E_A(\tocc{t_{h}}{a_{h}} \wedge \tocc{t'}{\trigg}).$$ 
	Choose $r'$ such that $(\Rrep,r')\sat \tocc{t_{h}}{a_{h}}$. Note that since 
	$\TTRi$ is solved in~$\Rrep$ and since $t_{h}-t_{g}=\delta_{h}-\delta_{g}$ 
	in every triggered run by problem definition, we get 
	$(\Rrep,r')\sat \tocc{t_{h}}{a_{h}}\leftrightarrow \tocc{t_{g}}{a_{g}}$.
	Since performing a local action is written, at least 
	for the current round, in the agent's local state (and hence known to the agent),  
	we obtain that $(\Rrep,r')\sat K_{\alpha_g}\tocc{t_{g}}{a_{g}}$, and hence
	also  $(\Rrep,r')\sat K_{\alpha_g}\tocc{t_{h}}{a_{h}}$.	
	Since $h$ is arbitrarily chosen in $\{1..k\}$, node $\alpha_{g}$ knows this for all responding nodes,
	and in particular for the earliest responding node $\alpha'=\node{i',t'}$. As responses always
	occur no sooner than the trigger, we get that  $(\Rrep,r')\sat K_{\alpha_g}\tocc{t'}{\trigg}$.
	Putting these results together we conclude 
	that $(\Rrep,r')\sat K_{i_g}(\tocc{t_{h}}{a_{h}}\wedge \tocc{t'}{\trigg})$. Since $g$ is arbitrarily 
	chosen in $\{1..k\}$, we get $(\Rrep,r')\sat E_A(\tocc{t_{h}}{a_{h}}\wedge \tocc{t'}{\trigg})$, from 
	which it follows that 
	$$(\Rrep,r')\sat \tocc{t_{h}}{a_{h}}\imp E_A(\tocc{t_{h}}{a_{h}}\wedge \tocc{t'}{\trigg})$$ 
	by our choice of $r'$. As false antecedents imply anything, we also get 
	$$(\Rrep,r'')\sat \tocc{t_{h}}{a_{h}}\imp E_A(\tocc{t_{h}}{a_{h}}\wedge \tocc{t'}{\trigg})$$ 
	in runs $r''$ where response $\alpha_{h}$ does not occur (non triggered runs) or where
	it occurs at a later time than $t_{h}$. We thus conclude that 
	$\Rrep\sat \tocc{t_{h}}{a_{h}}\imp E_A(\tocc{t_{h}}{a_{h}}\wedge \tocc{t'}{\trigg})$. 

    Recall the Knowledge Induction Rule in Lemma~\ref{lem:prop-nodeCK}, 
    that provides us with $\Rrep\sat \vphi\imp C_A\psi$ 
    from $\Rrep\sat \vphi\imp E_A(\vphi \wedge \psi)$. Setting $\vphi=\tocc{t_{h}}{a_{h}}$ 
    and $\psi=\tocc{t'}{\trigg}$ we apply the rule, and based on the above result obtain 
    $\Rrep\sat \tocc{t_{h}}{a_{h}}\imp C_A\tocc{t'}{\trigg}$. We conclude by taking notice that 
    $(\Rrep,r)\sat\tocc{t_{h}}{a_{h}}$ by assumption, and hence also $(\Rrep,r)\sat~C_A\tocc{t'}{\trigg}$.
\end{proofT}

\end{document}